\documentclass[aps,prl,twocolumn,groupedaddress,nofootinbib]{revtex4-1}

\usepackage{amsmath, amsfonts, amsthm, amssymb, graphicx}
\usepackage{xcolor}
\usepackage[utf8]{inputenc}
\usepackage{hyperref}

\usepackage[normalem]{ulem} 

\newcommand{\cmark}{ \checkmark }%
\newcommand{\xmark}{ \text{\sffamily X} }%

\def\be{\begin{equation}}
\def\ee{\end{equation}}
\def\ba{\begin{eqnarray}}
\def\ea{\end{eqnarray}}

\def\bq{\begin{quote}}
\def\eq{\end{quote}}

 at 10truept

\newcommand{\beq}{\begin{equation}}
\newcommand{\eeq}{\end{equation}}
\newcommand{\bea}{\begin{eqnarray}}
\newcommand{\eea}{\end{eqnarray}}
\newcommand{\beqa}{\begin{eqnarray}}
\newcommand{\eeqa}{\end{eqnarray}}

\newcommand{ \Stuckelberg}{St\"{u}ckelberg~}

\begin{document}
\raggedbottom

\title{UV Constraints on Massive Spinning Particles: 
Lessons from the Gravitino}

\author{Scott Melville} 
\email{scott.melville@damtp.cam.ac.uk}
\affiliation{DAMTP, Centre for Mathematical Sciences, University of Cambridge, CB3 0WA, United Kingdom}
\affiliation{Emmanuel College, University of Cambridge, CB2 3AP, United Kingdom}
\author{Diederik Roest} 
\email{d.roest@rug.nl}
\affiliation{Van Swinderen Institute for Particle Physics and Gravity, University of Groningen, \\ Nijenborgh 4, 9747 AG Groningen, The Netherlands}
\author{David Stefanyszyn} 
\email{d.stefanyszyn@damtp.cam.ac.uk}
\affiliation{DAMTP, Centre for Mathematical Sciences, University of Cambridge, CB3 0WA, United Kingdom}


\begin{abstract}
\noindent
Self-interacting massive particles with spin $\geq 1$ unavoidably violate unitarity; the question is at what scale. 
For spin-$1$ the strong coupling scale (at which perturbative unitarity is lost) cannot be raised by any finite tuning of the interactions, while for spin-$2$ 
there exists a special tuning of the Wilson coefficients which can raise this scale (and enjoys numerous special properties such as ghost-freedom). 
Here, we fill in the missing piece by describing how the self-interactions of a massive spin-$3/2$ 
field, or ``massive gravitino'', become strongly coupled at high energies. 
We show that while several different structures appear in the 
leading order potential, the strong coupling scale cannot be raised (in the absence of additional fields). 
At the level of the off-shell Lagrangian, it is always the non-linear symmetries of the longitudinal \Stuckelberg mode that dictate the strong coupling, and we show that in general it is only possible to parametrically raise the strong coupling scale if Wess-Zumino structures exist for these symmetries. 
We complement this off-shell approach with a first analysis of positivity bounds for a massive spin-$3/2$ particle, showing that any potential self-interaction which contributes to an on-shell 2-to-2 elastic process at tree level must vanish if this low-energy theory is to have a standard UV completion. We identify the mixing between the longitudinal mode and the transverse modes as the main obstacle to positivity, and clarify how the non-Abelian nature of non-linear (dRGT) massive gravity allows it to satisfy positivity where all known spin $\geq 3/2$ Abelian theories fail.  
Our results imply that a massive gravitino cannot appear alone in a controlled EFT---it must be accompanied by other particles, e.g.~as part of a supermultiplet. Together with the spin-$1$ and spin-$2$ cases, we suggest features which will persist in even higher spin massive theories. 
\end{abstract}

\maketitle

\noindent The low-energy effective field theory (EFT) description of a self-interacting massive vector or tensor field is well understood. 
The leading Lagrangians have a two-derivative kinetic sector that is gauge-invariant, plus a small mass term which softly breaks this gauge symmetry. 
For spin-$1$, this is the Proca action with a softly broken $U(1)$ gauge symmetry. 
For spin-$2$, the gauge symmetry can either be the full (non-Abelian) diffeomorphisms of General Relativity, with the kinetic sector given by the Einstein-Hilbert action, or the linearised (Abelian) version of diffeomorphisms, whose kinetic sector is that of linearised Einstein-Hilbert---the soft breaking of these symmetries gives rise to dRGT massive gravity \cite{MassiveGravity1, MassiveGravity2} and pseudolinear massive gravity \cite{pseudolinear1, pseudolinear2} respectively.  
Self-interactions in these theories are unavoidably non-renormalisable, and must be treated as an effective theory valid only below some cutoff (beyond which new degrees of freedom are required to restore unitarity\footnote{
For spin-$1$ the effective theory can be UV completed with the addition of a Higgs boson, while no Higgs mechanism (or any unitary UV completion) is known for the spin-$2$ case (see \cite{Bonifacio:2019mgk} and references therein). 
}). 

In this paper we fill in the gap between spin-$1$ and spin-$2$ by studying the EFT of a massive, self-interacting spin-$3/2$ 
field \cite{RaritaSchwinger}, which we shall refer to as a ``massive gravitino'' (although we are not considering supermultiplets). 
Such particles appear in many areas of physics---e.g. as QCD resonances, in supergravity and string theory---however a detailed study of their possible self-interactions is, until now, absent from the literature. 
We will approach the massive gravitino EFT from two complementary perspectives: an off-shell analysis of the strong coupling scales appearing in the Lagrangian (signalling when new physics must become important), and an on-shell analysis of the positivity properties of scattering amplitudes (revealing whether this new physics can be consistent with standard QFT axioms), drawing various parallels with the spin-$1$ and spin-$2$ cases as we proceed. By completing the catalogue of massive fields with spins $\leq 2$, we will draw some general conclusions/conjectures for arbitrary spin.

On the strong coupling side, if we assume a simple EFT power counting (naive dimensional analysis, or NDA \cite{Weinberg:1978kz, Manohar:1983md}) in which the gauge invariant sector becomes strongly coupled at the scale $\mathcal{M}_S$, and all symmetry breaking operators are universally suppressed by a single small parameter (which $\sim m^2_S/\mathcal{M}_S^2$ for bosons and $\sim m_S / \mathcal{M}_S$ for fermions), then the strong coupling scale\footnote{
Note that our estimated scale differs from that of \cite{PorratiRahman} since here we are considering self-interactions rather than coupling to an external electromagnetic field.
}  is naively 
 \begin{align}
 ( \mathcal{M}_S m_S^{3S-2} )^{\frac{1}{3S-1}} \,, \quad {\rm or} \quad ( \mathcal{M}_S m_S^{\frac{4S}{3} - 1} )^{\frac{3}{4S}} \,,
 \end{align} 
for a single spin-$S$ boson or fermion, respectively, of mass $m_S$. As $m_S \ll \mathcal{M}_S$, the symmetry breaking terms have lowered the cutoff substantially. 
For spin-$1$, no tuning of the Wilson coefficients can raise this strong coupling scale (except by turning off the interactions altogether). 
For spin-$2$, on the other hand, a particular $\mathcal{O}(1)$ tuning \emph{can} raise the strong coupling scale (from $(\mathcal{M}_S m_S^{4})^{1/5}$ to $(\mathcal{M}_S m_S^2)^{1/3}$). This special tuning also results in both dRGT and pseudolinear massive gravity being classically ghost-free at high energies \cite{MassiveGravity1,MassiveGravity2,pseudolinear1,pseudolinear2}. For higher integer spins raised strong coupling scales can be realised by the potentials of \cite{Bonifaciothesis,BrandoHS}.

For spin-$3/2$, while there are several different potential terms at leading order, we have found that no tuning of these operators can raise the strong coupling scale. 
In all of these cases, using the \Stuckelberg formulation of the off-shell Lagrangian (in which the broken gauge symmetry has been restored by means of additional fields), 
at high energies the onset of strong coupling is controlled by the longitudinal mode (which is a scalar for massive bosons and a spinor for massive fermions) and its non-linearly realised extended shift symmetries \cite{ExtendedShifts}. 
This allows us to connect (and go beyond) the spin-$1, 3/2$ and $2$ cases by showing that the ability to raise the strong coupling scale is tied to existence of Wess-Zumino structures for these shift symmetries:
 a result which readily generalises to all higher spin field theories.

We complement our off-shell approach with the first analysis of ``positivity bounds'' for the massive gravitino. 
These are constraints which the Wilson coefficients in the low-energy EFT must satisfy if it could have arisen from an underlying UV theory which is ``standard'' (in the sense of being unitary, causal/analytic, local and Lorentz invariant).
For spin-$1$ theories, there is clearly some region of parameter space which satisfies these bounds (since a massive vector could have arisen from a renormalisable Higgs mechanism in the UV). 
However, for spin-$2$ theories with an Abelian gauge symmetry (and with the NDA power counting), there is no region of parameter space which satisfies positivity \cite{pseudolinearpositivity}. 
We find that the massive gravitino, although its cutoff cannot be raised like the massive vector, shares the positivity fate of an Abelian spin-2: there is no interaction which contributes to the elastic 2-to-2 scattering amplitude (at tree-level) and satisfies positivity bounds. 
Unitary, causal/analytic, local and Lorentz invariant physics can never give rise to a single self-interacting massive spin-3/2 particle at low energies.

Across these different cases, it is the mixing between the longitudinal mode and the other transverse modes on which positivity hinges: for spin-$1$ this mixing does not grow fast enough with energy to affect positivity (and hence some region is allowed), while for spins $\geq 3/2$ this mixing appears in different amplitudes with opposite signs (requiring the whole vertex to be set to zero). 
This is further evidence for the observation made in \cite{BrandoHS} (with the aid of the spin-$3$ case) that Abelian higher spin theories in NDA are not consistent with positivity. 
Although positivity bounds can be satisfied in dRGT (non-Abelian) massive gravity \cite{MassiveGravityPositivity1, MassiveGravityPositivity2} (at sufficiently weak coupling \cite{deRham:2017xox, Bellazzini:2017fep}), we remark that it is the non-linear diffeomorphism invariance which is responsible for allowing particular two-derivative operators to enter at the same order as the potential interactions and rescue the positivity bounds which are violated in the Abelian case (without the non-linear symmetry, this tuning would violate the NDA power counting).  
  


We consider each spin-$S$ EFT in isolation, by which we mean that any other fields are sufficiently decoupled from the spin-$S$ field  that they provide only small corrections to its self-interactions. A possible resolution to the unitarity and positivity issues identified here would be to enlarge the particle content of the IR such that the spin-$S$ particle always couples to other light fields\footnote{
However, note that introducing couplings to new light degrees of freedom can also introduce new problems: for instance, when coupling to an electromagnetic field, charged particles with spin-$3/2$ famously suffer from the Velo-Zwanziger acausality problem \cite{VZ}, which can be resolved via specific non-minimal couplings \cite{PRRS}.
}. A natural possibility for spin-$3/2$ would be as part of a supermultiplet with a spin-$2$ field. 
Another possible resolution is to relax the NDA power counting, and allow derivative terms to compete with the potential interactions in the hope of partially cancelling the high-energy growth or rescuing the positivity bounds. We will return to this possibility in the Discussion.


\section{Strong Coupling and the \\ Goldstone Equivalence Theorem}
\noindent We begin the main body of the paper with a brief description of our EFT power counting scheme, followed by a review of strong coupling for massive spin-$1$ and spin-$2$ fields (with broken Abelian gauge symmetries) with a new understanding of why the strong coupling can be raised for spin-$2$ but not for spin-$1$. We then move onto our main focus of the massive gravitino.  \\

\noindent {\bf Power Counting - }
For a general spin-$S$ field, from a low energy EFT perspective the separation of scales between gauge invariant operators and those that break the gauge symmetry motivates the following power counting scheme, 
\begin{align}
\mathcal{L}_{\rm EFT} &=  \mathcal{M}_S^4 \, \mathcal{L}_{\rm gauge} \left(  \frac{\Phi}{ \mathcal{M}_S^{1+\theta/2} } \, , \, \frac{ \partial }{ \mathcal{M}_S }  \right) \nonumber \\ &+ m_S^{2-\theta} \mathcal{M}_S^{2+\theta} \,  V \left(\frac{ \Phi }{ \mathcal{M}_S ^{1+\theta /2}}  \, ,  \, \frac{ \partial }{ \mathcal{M}_S }  \right) \, , 
 \label{eqn:Vcounting}
\end{align}
where $\theta = 0,1$ for bosons and fermions respectively, $\mathcal{L}_{\rm gauge}$ is comprised of (dimensionless) gauge-invariant operators in the spin-$S$ field $\Phi$, while $V$ contains all (dimensionless) interactions which break the gauge symmetry.
From a UV perspective, $\mathcal{M}_S$ is related to the mass of a heavy field and $m_S/\mathcal{M}_S$ is related to a small symmetry-breaking parameter. From an IR perspective, $\mathcal{M}_S$ is the scale at which the gauge-invariant sector of the EFT breaks down, and $m_S \ll \mathcal{M}_S$ is the mass of the light particle $\Phi$. With this power counting it is the zero-derivative interactions which dominate any scattering amplitude.
%
For an $n$-boson (fermion) potential interaction, the naive strong coupling scale from \eqref{eqn:Vcounting} is $\Lambda_{\frac{nS+n-4}{n-2}}$ ($\Lambda_{ \frac{nS+n-4}{3n/2 - 3} } $), where $\Lambda_q = ( \mathcal{M}_S m_S^{q-1} )^{\frac{1}{q}} $, and is always lowest for the cubic (quartic) vertex and then increases monotonically. 
In the remainder of this section we explore under what conditions this strong coupling scale (and hence the maximum EFT cutoff) can be raised.
\\

\noindent {\bf Vector - } First consider a massive spin-$1$ field $A_{\mu}$ where the free Lagrangian is 
\begin{align} \label{Vector1}
\mathcal{L}_{S=1} = - \frac{1}{4}F_{\mu\nu}F^{\mu\nu} -\frac{1}{2}m_{1}^{2} A_{\mu}A^{\mu},
\end{align}
with\footnote{Here and throughout we (anti)-symmetrise with weight $n!$}
 $F_{\mu\nu} = \partial_{[\mu}A_{\nu]} = \partial_{\mu}A_{\nu} - \partial_{\nu}A_{\mu}$. The mass term breaks the $U(1)$ gauge symmetry $A_{\mu} \rightarrow A_{\mu} + \partial_{\mu} \Lambda(x)$ but this can be restored by making the \Stuckelberg replacement $A_{\mu} \rightarrow \hat{A}_{\mu} = A_{\mu} + \frac{1}{m_{1}} \partial_{\mu} \phi$. The resulting Lagrangian is invariant under $A_{\mu} \rightarrow A_{\mu} + \partial_{\mu} \Lambda(x)$, $\phi \rightarrow \phi - m_{1} \Lambda(x)$ and is given by
\begin{align} \label{Vector2}
\hat{\mathcal{L}}_{S=1} = \mathcal{L}_{S=1} -\frac{1}{2} (\partial \phi)^{2} - m_{1}A_{\mu} \partial^{\mu}\phi. 
\end{align}
In contrast to the unitary gauge Lagrangian \eqref{Vector1}, \eqref{Vector2} is smooth in the $m_{1} \rightarrow 0$ limit with three degrees of freedom both before and after taking the limit\footnote{Note that if we include a conserved source by augmenting the unitary gauge Lagrangian with $A_{\mu}J^{\mu}$, the longitudinal mode also decouples from the source in the massless limit.}.

In unitary gauge, the leading order quartic potential is simply,
\begin{align}
m_1^2 \mathcal{M}_1^2 \, V \left( \frac{A}{ \mathcal{M}_1 } \right) = \frac{m_{1}^{2}}{\mathcal{M}_{1}^{2}} \; C_1 \,  (A_{\mu}A^{\mu})^{2}  + ... ,
\label{Vspin1}
\end{align}
where $C_1$ is an (order unity) Wilson coefficient.  
After performing the \Stuckelberg replacement the potential takes the schematic form
\begin{align}
m_1^2 \mathcal{M}_1^2  V \left( \frac{A}{ \mathcal{M}_1 } \right) \sim \sum_{n=0}^{4}\frac{m_{1}^{2}}{\mathcal{M}_{1}^{2}} \frac{A^{n}(\partial \phi)^{4-n}}{m_{1}^{4-n}}.
\end{align}
The lowest scale suppressing these interactions is $\Lambda = \sqrt{ m_1 \mathcal{M}_{1} }$, corresponding to the $n=0$, dim-$8$ operator that only contains $(\partial \phi)$. 

This structure applies to any bosonic or fermionic massive higher spinning field: the quartic interactions that become strongly coupled first are the highest dimension operators which consist of four copies of the longitudinal mode each covered by $\lfloor S \rfloor$ derivatives. This follows straightforwardly from higher spin versions of the \Stuckelberg formulation, see e.g. \cite{MassiveGravityStuckelberg,PorratiRahman,BrandoHS}, and is simply the off-shell description of the Goldstone equivalence theorem (which has been extended to spin-3/2 fields \cite{GETFermion}). These highest dimension operators are invariant under a set of extended shift symmetries which for the scalar longitudinal mode are of the form
\begin{align} \label{extendedshifts}
\delta \phi = \sum_{n=0}^{S-1} a_{\mu_{1} \ldots \mu_{n}}x^{\mu_{1}} \ldots x^{\mu_{n}}.
\end{align}
For massive fermions, the symmetries of the spinor longitudinal mode are given by \eqref{extendedshifts} with the addition of a spinor index on the constant parameters $a_{\mu_{1} \ldots \mu_{n}}$. For spin-$1$ the relevant symmetry is simply a constant shift $\phi \rightarrow \phi + a$. 

Now to raise the strong coupling scale we require the highest dimension operators to form total derivative structures such that the corresponding vertices are trivial, even off-shell (such that they do not contribute to any scattering process). For spin-$1$ it is obvious that this is not possible since there is a unique and non-trivial dim-$8$ operator, but let us offer a complementary proof that no total derivative structure exists which can be easily generalised to higher spins. 

Without loss of generality, invariant dim-$8$ operators for a shift symmetric scalar can be written as $\partial_{\mu}\phi J^{\mu}$ where schematically $J \sim (\partial \phi)^{3}$. Now if this object is to only contribute a total derivative to the Lagrangian, we require $\partial_{\mu} J^{\mu} = 0$. This therefore constitutes an \textit{off-shell Noether current} corresponding to the shift symmetry (there is no need to go on-shell as we are classifying total derivatives). Its conservation is equivalent to the closure of the dual 3-form $H_3 = \star J_1$. Closure of $H_{3}$ ensures that  we can locally write $H_{3} = d B_{2} = d(\phi F_{2}) = d \phi \wedge F_{2} + \phi dF_{2}$ where $F_{2}$ is an invariant $2$-form with two derivatives and two fields. Moreover, invariance of $H_{3}$ w.r.t.~the constant shift symmetry requires $F_2$ to be closed. Again we have locally $F_{2} = d(\phi G_{1}) = d \phi \wedge G_{1} + \phi dG_{1}$ where $G_{1}$ has a single derivative and a single field. Invariance of $F_{2}$ in turn requires closure of $G_{1}$ and hence $G_{1} = C_0 d \phi$ for an arbitrary constant $C_0$. It is then clear that no cubic off-shell current  (and therefore also no quartic total derivative) exists since $F_{2} = C_0 d \phi \wedge d \phi = 0$.  

The absence of a total derivative structure, and therefore the absence of a raised strong coupling scale, is directly linked to absence of an \textit{interacting} WZ term for the scalar's shift symmetry. WZ operators differ from strictly invariant operators (which are simply built out of $\partial_{\mu} \phi$) since $i)$ they are only invariant up to a total derivative and $ii)$ they have a different power counting with fewer derivatives per field. The absence of a WZ term in this case is easy to see via the coset construction \cite{InternalCoset1,InternalCoset2,SpacetimeCoset} (see e.g.~\cite{GalileonWessZuminos, KRS} for reviews) with symmetry breaking pattern 
\begin{align}
\frac{G}{H} = \frac{ISO(1,3) \times U(1)}{SO(1,3)}.
\end{align}
Indeed a shift symmetric scalar is the Goldstone boson of a spontaneously broken $U(1)$ symmetry. Parametrising the coset element as 
\begin{align}
\Omega = e^{x^{\mu}P_{\mu}}e^{\phi Q},
\end{align}
yields the Maurer-Cartan (MC) 1-form
\begin{align}
\Omega^{-1} d \Omega = dx^{\mu}P_{\mu} + d \phi Q.
\end{align}
In 4D a WZ term comes from constructing a closed $5$-form $\beta_{5} = d \beta_{4}$ out of the MC $1$-forms, followed by pulling $\beta_{4}$  back to spacetime. Clearly, any $5$-form with four copies of the scalar will include the vanishing $2$-form $F_{2} = d \phi \wedge d\phi$. So whether we are trying to construct a dim-$8$ total derivative or an interacting WZ term, boils down to performing the same computations of off-shell currents. 

In conclusion, our inability to raise the strong coupling scale for a massive spin-$1$ field is directly related to the absence of an interacting\footnote{A linear potential is a WZ term and arises from the 5-form $\beta_{5} = \epsilon_{\mu\nu\rho\sigma} dx^{\mu} \wedge dx^{\nu} \wedge dx^{\rho} \wedge dx^{\sigma} \wedge d \phi$.} WZ structure for the shift symmetric longitudinal mode.  \\ 


\noindent
{\bf Tensor - }
The analysis for a massive spin-$2$ field $h_{\mu\nu}$ with a broken Abelian gauge symmetry proceeds along similar lines but with qualitatively different results. 
 Here the free Lagrangian is given by
 \begin{align}
\mathcal{L}_{S=2} = -\frac{1}{2} h_\alpha^{\; [\alpha} \partial_\beta \partial^{\beta} h_{\gamma}^{\; \gamma]}  - \frac{1}{2}m_{2}^{2} ([h^{2}] - [h]^{2}),
\end{align}
where square brackets denote a trace with indices raised by $\eta^{\mu\nu}$ e.g. $[h] = \eta^{\mu\nu}h_{\mu\nu}$, $[h^{2}] = \eta^{\mu\nu}\eta^{\rho\sigma}h_{\mu\rho}h_{\nu\sigma}$. The kinetic sector is simply that of linearised General Relativity (up to total derivatives) while the Fierz-Pauli tuning between the mass terms ensures only five dynamical degrees of freedom \cite{FierzPauli}. The mass terms breaks the Abelian gauge symmetry $h_{\mu\nu} \rightarrow h_{\mu\nu} + \partial_{(\mu}\xi_{\nu)}$ but this can be restored by means of the \Stuckelberg replacement,
\begin{align} \label{GravitonStuckelberg}
h_{\mu\nu} \rightarrow \hat{h}_{\mu\nu} = h_{\mu\nu} + \frac{1}{m_{2}} \partial_{(\mu}A_{\nu)} + \frac{2}{m_{2}^{2}} \partial_{\mu}\partial_{\nu}\phi. 
\end{align}
The resulting Lagrangian is gauge invariant and includes all three fields in the massless limit. However, in contrast to the spin-$1$ case, the resulting kinetic sector is not yet diagonal and the field redefinition\footnote{In contrast to spin-$1$, the scalar does not decouple from a traceful, conserved source after this field redefinition thereby leading to the well-known vDVZ discontinuity \cite{vdvz1,vdvz2}.}
\begin{align} \label{GravitonFieldRedefinition}
h_{\mu\nu} \rightarrow \tilde{h}_{\mu\nu} = h_{\mu\nu} + \eta_{\mu\nu}\phi,
\end{align}
is required to kinetically decouple the different helicity modes such that each propagator has the desired high energy behaviour. We refer the reader to e.g. \cite{MassiveGravityReview1} for explicit forms of the Lagrangian at each step.

Unlike the spin-1 case, the symmetry-breaking potential now contains a number of independent terms at each order in the field, beginning with three cubic operators, $c_1 [ h^3] + c_2 [h] [h^2] + c_3 [h]^3$, and followed by five quartic operators,
\begin{align}
m_2^2 \mathcal{M}_2^2 \, V \left( \frac{h}{ \mathcal{M}_2 } \right) = \frac{ m_{2}^{2} }{\mathcal{M}_{2}^{2}} ( &d_{1}  [h^{4}] + d_{2} [h^{3}][h] + d_{3}[h^{2}]^{2} +  \nonumber \\ &+ d_{4}[h^{2}][h]^{2} + d_{5}[h]^{4}),
\label{eqn:spin2quartics}
\end{align}
where the dimensionless constants $d_{i}$ are all $\mathcal{O} (1)$ or smaller. 
We will focus on the quartic potential to facilitate comparison with the spin-$1$ and spin-$3/2$ theories (which have only quartic terms in their respective potentials).
By performing the replacements \eqref{GravitonStuckelberg} and \eqref{GravitonFieldRedefinition} we find various interactions of the schematic form 
\begin{align}
m_2^2 \mathcal{M}_2^2 \, &V \left( \frac{h}{ \mathcal{M}_2 } \right) \sim \nonumber \\ &\sum_{m=0}^{4} \sum_{n=0}^{4-m}\frac{m_{2}^{2}}{\mathcal{M}_{2}^{2}} \frac{(h + \eta \phi)^{m} (\partial A)^{n}(\partial \partial \phi)^{4-n-m}}{m_{2}^{8-2m-n}},
\end{align}
with operators ranging from dim-$4$ through to dim-$12$. 
The dim-$12$ interactions, of the schematic form $(\partial \partial \phi)^{4}$, become strongly coupled first at $\Lambda = ( m_2^3 \mathcal{M}_{2} )^{1/4}$.

To raise the strong coupling scale we therefore need to make these vertices off-shell trivial i.e. turn them into total derivatives by tuning the $d_{i}$. These operators are trivially invariant under the global non-linear symmetries $\phi \rightarrow \phi + a + b_{\mu}x^{\mu}$ given that each copy of the scalar is covered by two derivatives. These are the symmetries of the Galileon \cite{Galileon}. Any quartic operator with these symmetries can be written as $\partial_{\mu}\partial_{\nu}\phi J^{\mu\nu}$ where $J \sim (\partial \partial \phi)^{3}$; again, if these are to form a total derivative structure we require $\partial_{\mu} J^{\mu\nu} = 0$ \textit{off-shell}. Off-shell conservation implies closure of the dual vector-valued 3-form $H_3^\mu = \star J_1^\mu$ (where we interpret the current as a vector valued 1-form $J_1^\mu = J^\mu{}_{\nu} dx^\nu$), and hence $H_3^\mu = d ( \partial_\nu \phi F_2^{\mu \nu} )$. 
Invariance of $H_3$ implies closure of $F^{\mu \nu}_2 = d (\partial_\rho \phi G_1^{\mu \nu \rho})$. In this case we find that $G_1^{\mu \nu \rho} = C_0^{\mu \nu \rho \sigma} d (\partial_\sigma \phi)$, where $C$ is a four-index Lorentz invariant tensor. The two options are the Levi-Civita tensor or the product of two metrics. 
The former gives rise to an invariant and closed $H_3^\mu = \epsilon^{\mu \nu \rho \sigma} (d \partial_\nu \phi) \wedge (d \partial_\rho \phi) \wedge (d \partial_\sigma \phi)$ and therefore a single total derivative structure. It can easily be seen that the alternative option with two metrics leads to a vanishing 3-form.
 
We therefore see that it is indeed possible to tune the $d_{i}$ such that the dim-$12$ operators combine into a total derivative and this requires the unitary gauge potential to take the form\cite{MassiveGravity1}
\begin{align}
m_2^2 \mathcal{M}_2^2 \, V \left( \frac{h}{ \mathcal{M}_2 } \right) = \frac{m_{2}^{2}}{\mathcal{M}_{2}^{2}}\epsilon^{\mu_{1}..\mu_{4}}\epsilon^{\nu_{1}..\nu_{4}}h_{\mu_{1}\nu_{1}} h_{\mu_{2}\nu_{2}} h_{\mu_{3}\nu_{3}}h_{\mu_{4}\nu_{4}}.
\label{eqn:hhhh}
\end{align}
It is now the dim-$10$ operators of the schematic form $\phi (\partial \partial \phi)^{3}$ and $(\partial A)^{2} (\partial \partial \phi)^{2}$ that become strongly coupled first (the dim-$11$ operator is also a total derivative). The new strong coupling scale is $\Lambda_{*} =  ( m_2^2 \mathcal{M}_2 )^{1/3}$, often referred to as $\Lambda_3$, and is $\gg \Lambda$ without tuning. For more details see the reviews \cite{MassiveGravityReview1, MassiveGravityReview2}. 

We will now show that our ability to raise the strong coupling scale is directly linked to the presence of an interacting WZ term for the non-linear symmetries of the longitudinal mode. The algebra of these symmetries and the linearly realised Poincar\'{e} symmetries forms the five-dimensional Galileon algebra which is a contraction of $ISO(1,4)$. The derivation of invariants and WZ terms follows in the same manner as above with the addition of inverse Higgs constraints that can appear when a spacetime symmetry is spontaneously broken \cite{InverseHiggs, LowInverseHiggs}. Since here the spontaneously broken generators are one scalar and one vector, one would naively expect both a scalar and vector field in the resulting non-linear realisation. However, Goldstone's theorem \cite{GoldstoneTheorem} does not apply beyond the breaking of internal symmetries and indeed the Galileon algebra is a spacetime symmetry (since it is not a direct product $ISO(1,3) \times ...$). An inverse Higgs constraint exists that eliminates the would-be vector mode in favour of the first derivative of the massless scalar\footnote{Another way of seeing that this vector cannot be an integral part of the low energy realisation is that it is gapped and can therefore be integrated out of the path integral at energies below its mass.}. The details of the inverse Higgs phenomenon within this context and the WZ terms described below was worked out in \cite{GalileonWessZuminos} and we refer the reader there for more details.

The only non-trivial commutator that involves the broken generators $A$ and $B_{\mu}$ (corresponding to symmetry parameters $a$ and $b_{\mu}$) is $[P_{\mu},B_{\nu}] = \eta_{\mu\nu}A$. The other commutators are those of the linearly realised $ISO(1,3)$ sub-algebra and those that define the Lorentz representation of $A$ and $B_{\mu}$. Parametrising the coset element as 
\begin{align}
\Omega = e^{x^{\mu}P_{\mu}}e^{\phi A}e^{\phi_{\mu}B^{\mu}},
\end{align} 
yields the MC form
\begin{align}
\Omega^{-1}d \Omega = dx^{\mu}P_{\mu} + (d \phi + \phi_{\mu}dx^{\mu})A + d \phi^{\mu}B_{\mu}.
\end{align}
To solve for $\phi_{\mu}$ we set to zero the MC form along the broken generator $A$ yielding $\phi_{\mu} = - \partial_{\mu} \phi$ from which we see that invariant Lagrangians are constructed out of $\partial_{\mu}\partial_{\nu}\phi$ once we pullback to 4D spacetime\footnote{Inverse Higgs constraints actually correspond to setting to zero covariant derivatives which are the product of the coset vielbein and the corresponding MC component. However, for extended shift symmetries the vielbein is the identity and so if the covariant derivative is a Lorentz irrep (as it is here), setting the MC form itself to zero is equivalent.}.

Turning to interacting WZ terms (the scalar's kinetic term is also a WZ \cite{GalileonWessZuminos}), any four-point vertices must come from a $5$-form linear in $dx^{\mu}$ with the remaining four $1$-forms coming from the components along $A$ and $B_{\mu}$ which we respectively denote as $\omega_{A}$ and $\omega_{B}^{\mu}$. Now, in the presence of inverse Higgs constraints one constructs closed $5$-forms, pulls the resulting $4$-form back to spacetime followed by imposing the inverse Higgs constraint. However, it is clear from this procedure that any $5$-form non-linear in an object that vanishes on the inverse Higgs solution, in this case $\omega_{A}$, will not yield a non-trivial WZ term. The $5$-form will therefore have to be linear in $dx^{\mu}$, linear in $\omega_{A}$ and cubic in $\omega_{B}$, of which there is only one possibility (due to symmetry), 
\begin{align}
\beta_{5} = \epsilon_{\mu\nu\rho\sigma} dx^{\mu} \wedge \omega_{A} \wedge \omega_{B}^{\nu} \wedge \omega_{B}^{\rho} \wedge \omega_{B}^{\sigma},
\end{align}
which is indeed closed \cite{GalileonWessZuminos}. After peeling off the $ dx^{\mu} \wedge \omega_{A}$ part, this is exactly the closed and invariant 3-form that we constructed before and the corresponding WZ interaction is the quartic Galileon. As in the vector case, we therefore find a relation between the existence of a quartic total derivative and the corresponding WZ interaction. 
For the massive tensor, there is also a cubic (and quintic) WZ term, reflecting the fact that an analogous tuning is possible for the cubic (and quintic) potential, which raises the strong coupling scale from the naive estimate of $( \mathcal{M}_2 m_2^4 )^{1/5}$ (or $ ( \mathcal{M}_2 m_2^{8/3} )^{3/11}  $) to $(  \mathcal{M}_2 m_2^2 )^{1/3}$.   

Note that within the NDA power counting \eqref{eqn:Vcounting}, an Abelian symmetry guarantees that the kinetic term in $\mathcal{L}_{\rm gauge}$ is only quadratic in the fields and does not produce interactions---two-derivative operators like $\partial^2 h^3$ and $\partial^2 h^4$ are therefore viewed as small symmetry-breaking interactions and are suppressed relative to the zero-derivative potential considered above (by a factor of $\partial^2/\mathcal{M}_2^2$). 
If one relaxes this power counting to allow for ``anomalously large'' two-derivative couplings, whose Wilson coefficients are enhanced (by a factor of $\mathcal{M}_2^2/m_2^2$) so that they contribute at the same order as the zero-derivative potential (since $\partial^2/m_2^2 \sim 1$), then it becomes possible to raise the cutoff in two distinct ways. 
One possibility is to retain the tuning \eqref{eqn:hhhh} described above (ensuring that the zero-derivative potentials become strongly coupled at $\Lambda_3$), and fix the coefficients of the two-derivative terms to independently set their strong coupling scale to $\Lambda_3$ as well---this was performed in \cite{pseudolinear2} and gives the two-derivative interaction of pseudolinear massive gravity. But a second possibility is to search for a tuning which cancels the high energy growth of the zero-derivative potential using the two-derivative interactions---one solution\footnote{
The results of \cite{BonifacioHinterbichler1} suggest that this is the unique solution beyond the Abelian pseudolinear tuning.
} being the perturbative expansion of $\sqrt{-g} R = \partial^2 h^2 + \partial^2 h^3 + ... $ about a flat background ($g_{\mu\nu} = \eta_{\mu \nu} + h_{\mu \nu}$) and the corresponding $c_i$, $d_i$, etc. tunings of dRGT massive gravity \cite{MassiveGravity1}. The merit of this solution is that the ``violation'' of NDA (the ``large'' values of the two-derivative coupling constants) is actually protected by a new, non-Abelian, symmetry: non-linear diffeomorphisms. The EFT is still in the radiatively stable form \eqref{eqn:Vcounting}, but has reorganised so that $\mathcal{L}_{\rm gauge}$ now contains all interactions invariant under the non-Abelian symmetry, and consequently its interactions (e.g. $\partial^2 h^3/M_P$) are now naturally the same order as zero-derivative terms in the potential (e.g. $m^2 h^3/M_P$). 
In this non-Abelian case, since the power counting \eqref{eqn:Vcounting} still applies, the connection between WZ structures and the raising of the cutoff also still holds. Rather than the linear replacement \eqref{GravitonStuckelberg}, the correct \Stuckelberg procedure for the non-Abelian symmetry is described in \cite{MassiveGravity2}, and the dRGT tuning which raises the cutoff corresponds to ensuring that the longitudinal mode has interactions of the form $\partial_\mu \partial_\nu \phi J^{\mu \nu}$, where $J^{\mu \nu}$ is conserved off-shell (as described above).   
\\      



\noindent
{\bf Gravitino - }
We now turn to spin-$3/2$ which is the main focus of this work. Although the free Lagrangian for a massive gravitino is well-known \cite{RaritaSchwinger}, the quartic self-interactions are not and finding a complete basis is made somewhat complicated by Fierz identities. However, we can avoid this issue by employing $SU(2) \times SU(2)$ notation for all indices by converting vectors to bispinors via $v_{\alpha \dot{\alpha}} = (\sigma^{\mu})_{\alpha \dot{\alpha}}v_{\mu}$. A vector-spinor is therefore represented by the Weyl spinor $\psi_{\alpha \beta \dot{\alpha}}$ and its complex conjugate $(\psi_{\alpha \beta \dot{\alpha}})^{\dagger} = \bar{\psi}_{\alpha \dot{\alpha}\dot{\beta}}$. Indices are raised and lowered with the epsilon tensors $\epsilon_{\alpha \beta}, \epsilon_{\dot{\alpha}\dot{\beta}}$ e.g. $\epsilon^{\alpha \beta}\psi_{\beta \alpha \dot{\alpha}} = \psi^{\alpha}{}_{\alpha \dot{\alpha}} $ and  throughout we follow the conventions of \cite{Conventions}. When we do introduce a mixture of indices, we will use the start of the Greek alphabet ($\alpha, \beta, \ldots$) for spinor indices and the middle ($\mu,\nu,\ldots$) for Lorentz indices. 

The free Lagrangian is
\begin{align} \label{RS}
&\mathcal{L}_{S=3/2} = i \bar{\psi}^{\alpha}{}_{(\dot{\alpha} \dot{\beta})} \partial^{\beta \dot{\alpha}}\psi_{(\alpha \beta)}{}^{\dot{\beta}} + i \bar{\psi}^{\alpha}{}_{\dot{\alpha}}{}^{\dot{\alpha}} \partial^{\beta}{}_{\dot{\beta}} \psi_{(\alpha \beta)}{}^{\dot{\beta}}  \nonumber \\&- i \bar{\psi}^{\gamma}{}_{(\dot{\beta}\dot{\gamma})} \partial_{\gamma}{}^{\dot{\gamma}} \psi^{\alpha}{}_{\alpha}{}^{\dot{\beta}} \nonumber + 3i \bar{\psi}^{\alpha}{}_{\dot{\alpha}}{}^{\dot{\alpha}} \partial_{\alpha \dot{\beta}}\psi^{\gamma}{}_{\gamma}{}^{\dot{\beta}} \\ & - m_{3/2} \left(\psi^{(\alpha \beta)}{}_{\dot{\alpha}}\psi_{\alpha \beta}{}^{\dot{\alpha}} - 3 \psi^{\alpha}{}_{\alpha \dot{\alpha}}\psi^{\beta}{}_{\beta}{}^{\dot{\alpha}} + c.c \right),
\end{align}
where we have decomposed the field into two irreps: a traceless part $\psi_{(\alpha \beta) \dot{\alpha}} = \psi_{\alpha \beta \dot{\alpha}} + \psi_{\beta \alpha \dot{\alpha}}$ and a pure trace $\psi^{\alpha}{}_{\alpha \dot{\alpha}}$. We remind the reader that since traces are taken with the anti-symmetric epsilon tensors, sets of symmetric indices are traceless and therefore irreducible. The tuning in the kinetic sector ensures that this part of the action is invariant under the Abelian gauge symmetry
\begin{align}
\psi_{\alpha \beta \dot{\alpha}} \rightarrow \psi_{\alpha \beta \dot{\alpha}} + \partial_{\alpha \dot{\alpha}} \epsilon_{\beta} \label{FermionGaugeSymm},
\end{align}
where $\epsilon = \epsilon(x)$ is a fermionic gauge parameter, while the tuning between the mass terms ensures that the longitudinal spinor is not a ghost (i.e.~it has a single derivative kinetic term) and hence there are four degrees of freedom.

Now as with the bosonic cases above, this action does not admit a smooth $m_{3/2} \rightarrow 0$ limit since the mass term breaks the gauge symmetry of the kinetic sector. To remedy this and ensure a smooth massless limit, we now restore the broken gauge symmetry by introducing a \Stuckelberg field $\lambda_{\alpha}$. By sending
\begin{align}
&\psi_{\alpha \beta \dot{\alpha}} \rightarrow \hat{\psi}_{\alpha \beta \dot{\alpha}} = \psi_{\alpha \beta \dot{\alpha}}  + \frac{1}{m_{3/2}} \partial_{ \alpha \dot{\alpha}} \lambda_{\beta},
\end{align}
the action is manifestly invariant under \eqref{FermionGaugeSymm}  with $\lambda_{\alpha} \rightarrow \lambda_{\alpha} - m_{3/2} \epsilon_{\alpha}$. Under this \Stuckelberg replacement \eqref{RS} becomes
\begin{align} \label{RS1}
&\hat{\mathcal{L}}_{S=3/2} = \mathcal{L}_{S=3/2} - (2 \psi^{(\alpha \beta)}{}_{\dot{\alpha}} \partial_{\alpha}{}^{\dot{\alpha}}\lambda_{\beta} - 6 \psi^{\alpha}{}_{\alpha \dot{\alpha}} \partial^{\beta \dot{\alpha}}\lambda_{\beta} +  c.c).
 \end{align}
As in the spin-$2$ case, we now to need diagonalise the kinetic terms in \eqref{RS1} to ensure well-behaved propagators at high energies. Using the following shift in the pure trace 
\begin{align} \label{RSFieldRedefinition}
\psi^{\alpha}{}_{\alpha \dot{\alpha}} \rightarrow \tilde{\psi}^{\alpha}{}_{\alpha \dot{\alpha}}  = \psi^{\alpha}{}_{\alpha \dot{\alpha}} + 2i \bar{\lambda}_{\dot{\alpha}},
\end{align}
we find
\begin{align} \label{RS2}
\tilde{\mathcal{L}}_{S=3/2} =& \mathcal{L}_{S=3/2} + 12i \bar{\lambda}_{\dot{\alpha}} \partial^{\alpha \dot{\alpha}} \lambda_{\alpha} \nonumber \\ &+12m_{3/2} (i \bar{\lambda}_{\dot{\alpha}} \psi^{\beta}{}_{\beta}{}^{\dot{\alpha}} - \bar{\lambda}_{\dot{\alpha}} \bar{\lambda}^{\dot{\alpha}} +  c.c),
 \end{align}
and now when we send $m_{3/2} \rightarrow 0$ we are left with a massless spin-$3/2$ mode and the massless spinor longitudinal mode, as required\footnote{As is the case for spin-$2$, the longitudinal mode does not decouple from a traceful, conserved source.}.
The same procedure using four-component spinors can be found in \cite{PorratiRahman} (see also \cite{SUSYUM} for a superspace version of the \Stuckelberg formulation).

We now switch on a unitary gauge quartic potential for the massive gravitino and analyse the strong coupling. To construct a complete basis we simply write down all possible Lorentz scalars, organised in terms of the number of traces, followed by making repeated use of identities like
\begin{align}
\psi_{\beta}{}^{\alpha}{}_{\dot{\alpha}} \psi_{\gamma \alpha}{}^{\dot{\alpha}} = \frac{1}{2} \epsilon_{\beta \gamma} \psi^{\delta \alpha}{}_{\dot{\alpha}} \psi_{\delta \alpha}{}^{\dot{\alpha}},
\end{align}
to identify degeneracies. In the end we find eight linearly independent operators, two of the form $\psi^{4} + c.c$ and six of the form $\psi^{2}\bar{\psi}^{2}$. To write the interactions in a compact form we introduce the traces $\psi^{\alpha}{}_{\alpha \dot{\alpha}} = [\psi]_{\dot{\alpha}}$, $\bar{\psi}_{\dot{\alpha}}{}^{\dot{\alpha} \alpha} = [\bar{\psi}]^{\alpha}$ and define the following scalar and bi-spinor products
\begin{align}
\psi \cdot \psi &= \psi^{(\alpha \beta)}{}_{\dot{\alpha}}\psi_{\alpha \beta}{}^{\dot{\alpha}} \\ 
\bar{\psi} \cdot \bar{\psi} &= \bar{\psi}^{\alpha}{}_{(\dot{\alpha} \dot{\beta})}\bar{\psi}_{\alpha}{}^{\dot{\alpha} \dot{\beta}} \\ (\psi \cdot \bar{\psi})^{\alpha \dot{\alpha}} &= \psi^{(\alpha \beta)}{}_{\dot{\beta}} \bar{\psi}^{(\dot{\beta} \dot{\alpha})}{}_{\beta}.
\end{align}
The leading order quartic potential is then\footnote{See appendix A for a complete basis in the more familiar notation using a mixture of $SO(1,3)$ and $SU(2) \times SU(2)$ indices.}
\begin{align} \label{quartics}
m_{3/2} \mathcal{M}_{3/2}^3  & V \left( \frac{ \psi }{\mathcal{M}_{3/2}^{3/2} }  \right)    \nonumber \\
 =  \frac{ m_{3/2}  }{ \mathcal{M}_{3/2}^3 } & \Bigg[ (g_{1}[\psi] \cdot [\psi] \psi \cdot \psi + g_{2} (\psi \cdot \psi)^{2} + c.c) \nonumber \\
&+g_{3} [\psi] \cdot [\psi] [\bar{\psi}]\cdot [\bar{\psi}] +g_{4} (\psi \cdot \bar{\psi}) \cdot [\psi] \cdot [\bar{\psi}] \nonumber \\
&+(g_{5} [\psi] \cdot [\psi] \bar{\psi} \cdot \bar{\psi} + c.c) \nonumber \\
&+(g_{6} (\psi \cdot \bar{\psi}) \cdot ([\psi] \cdot \bar{\psi}) + c.c) \nonumber \\
&+g_{7} (\psi \cdot \psi) (\bar{\psi} \cdot \bar{\psi}) + g_{8} (\psi \cdot \bar{\psi}) \cdot (\psi \cdot \bar{\psi})   \Bigg] ,
\end{align} 
where each real dimensionless coefficient is order unity, $g_{i} \sim \mathcal{O}(1)$. 

Now when we perform the \Stuckelberg replacement and diagonalise the kinetic terms, the potential takes the schematic form
\begin{align}
m_{3/2} \mathcal{M}_{3/2}^3  & V \left( \frac{ \psi }{\mathcal{M}_{3/2}^{3/2} }  \right) \sim \sum_{n=0}^{4}\frac{m_{3/2} }{\mathcal{M}_{3/2}^{3}} \frac{(\psi + \bar{\lambda})^{n} (\partial \lambda)^{4-n}}{m_{3/2}^{4-n}},
\end{align}
with dim-$6$ through to dim-$10$ operators. 
As expected it is the dim-$10$ operators of the form $( \partial \lambda)^{4}$ that become strongly coupled first, at $\sqrt{m_{3/2} \mathcal{M}_{3/2}}$. To raise the strong coupling scale these operators must combine into a total derivative thereby rendering them off-shell trivial. Note that in contrast to the spin-$1$ case, there are a number of different non-trivial dim-$10$ operators so there is at least in principle a possibility of finding a total derivative structure by tuning $g_{i}$.

These operators are invariant under the constant shift symmetry $\lambda^{\alpha} \rightarrow \lambda^{\alpha} + \chi^{\alpha}$ and any invariant dim-$10$ operator can be written as $\partial_{\mu}\lambda_{\alpha}J^{\mu \alpha} + c.c$ where schematically $J \sim (\partial \lambda)^{3}$. The existence of a total derivative then requires $\partial_{\mu} J^{\mu \alpha} = 0$; it is the off-shell conserved current associated with the fermionic shift symmetry. Its conservation implies closure of the dual, spinor-valued 3-form $H_{3}^{\alpha}$ with three derivatives and three fields (barred or unbarred). Locally we can write $H_{3}^{\alpha} = d(\lambda_{\beta}F^{\alpha \beta}_{2} + \bar{\lambda}_{\dot{\beta}}\tilde{F}_{2}^{\alpha \dot{\beta}})$ where $F_{2}$ and $\tilde{F}_{2}$ have two derivatives and two fields. Invariance of $H_{3}$ with respect to the fermionic shift symmetry implies the closure of both $F_2$ and $\tilde F_2$. This implies $F_{2}^{\alpha \beta}  = d(\lambda_{\gamma} G_{1}^{\alpha \beta \gamma} + \bar{\lambda}_{\dot{\gamma}} \tilde{G}_{1}^{\alpha \beta \dot{\gamma}})$, and a similar structure for $\tilde{F}_{2}$. Invariance of the $F's$ requires the 1-forms to be closed, and therefore to take e.g.~the form $G_1^{\alpha \beta \gamma} = C_0^{\alpha \beta \gamma \delta} d (\lambda_\delta )$ with $C_0$ in terms of invariant Levi-Civita tensors. Putting things together, the spinor-valued $3$-form $H_{3}^{\alpha}$ is constructed out of the $1$-forms $d \lambda^{\alpha}$, $d \bar{\lambda}^{\dot{\alpha}}$, which however will always vanish as  $d \lambda^{\alpha} \wedge d \lambda_{\alpha} = d\bar{\lambda}_{\dot{\alpha}} \wedge d\bar{\lambda}^{\dot{\alpha}} = 0$. We therefore conclude that no tunings of the coefficients $g_{i}$ can turn the dim-$10$ operators in a total derivative and therefore there is no way to raise the strong coupling scale (see appendix A for a proof of the absence of a total derivative structure directly at the level of the equations of motion for the longitudinal spinor).

We now look for interacting WZ terms for the fermion's constant shift symmetry. The shift symmetry of course commutes with itself and translations, but it does not commute with the Lorentz generators since it transforms in the $(1/2,0) + (0,1/2)$ representation of the Lorentz group. This shift symmetry is the first possible extended shift symmetry for a spinor. We denote the generators of this symmetry as $Q_{\alpha}$ and $\bar{Q}_{\dot{\alpha}}$. The relevant algebra is a contraction of $\mathcal{N} =1$ super-Poincar\'{e} \cite{InternalSUSY} (incidentally, the super-Poincar\'{e} algebra is the only non-Abelian algebra that can be non-linearly realised by a single Weyl fermion \cite{RSWexceptional}). We parametrise the coset element as
\begin{align}
\Omega = e^{x^{\mu}P_{\mu}}e^{\lambda^{\alpha} Q_{\alpha} + \bar{\lambda}_{\dot{\alpha}}\bar{Q}^{\dot{\alpha}}},
\end{align}
yielding the MC form 
\begin{align}
\Omega^{-1}d \Omega = dx^{\mu}P_{\mu} + d \lambda^{\alpha}Q_{\alpha} + d \bar{\lambda}_{\dot{\alpha}} \bar{Q}^{\dot{\alpha}}.
\end{align}
It follows that strictly invariant Lagrangians are constructed out of $\partial_{\mu}\lambda^{\alpha}$ and $\partial_{\mu}\bar{\lambda}_{\dot{\alpha}}$ and so even the Weyl kinetic term, which has only a single derivative, must be a WZ term and indeed it is. The relevant $5$-form is 
\begin{align}
\beta_{5} = \epsilon_{\mu\nu\rho\kappa}\sigma^{\kappa}_{\alpha \dot{\alpha}} dx^{\mu} \wedge dx^{\nu} \wedge dx^{\rho} \wedge d \lambda^{\alpha} \wedge d \bar{\lambda}^{\dot{\alpha}}.
\end{align}
If a WZ quartic interaction is to exist the corresponding $5$-form must be of the form
\begin{align}
\beta_{5} = (\sigma_{\mu})_{\alpha \dot{\alpha}} dx^{\mu} \wedge d \bar{\lambda}^{\dot{\alpha}} \wedge H_{3}^{\alpha}
\end{align}
where $H_{3}^{\alpha}$ is a spinor-valued $3$-form constructed out of $d \lambda^{\alpha}$ and $d \bar{\lambda}^{\dot{\alpha}}$. We already know from our search for dim-$10$ total derivatives that such an object does not exist. We therefore see again that our inability to raise the strong coupling scale for the massive gravitino is directly related to the absence of an interacting WZ term for the longitudinal spinor's constant shift symmetry.

\section{Positivity and the \\ Nature of UV Completion}

\noindent We have established that the strong coupling scale of a massive gravitino cannot be raised in the same way as a massive graviton. Moreover, it has been shown that theories of a massive spin-2 particle in the IR can only arise from ``consistent'' UV physics (which is unitary, analytic, local and Lorentz invariant) if $i)$ the broken gauge symmetry is non-Abelian i.e. the kinetic sector is given by the full Einstein-Hilbert action and $ii)$ the potential is tuned in the way which raises its strong coupling scale to $\Lambda_{*} $\cite{MassiveGravityPositivity2} i.e. it is of the dRGT form (for two interacting spin-2 fields positivity bounds also favour the EFT with the highest strong coupling scale \cite{Alberte:2019xfh}). The fact that the pseudolinear theory is inconsistent with positivity bounds was shown in \cite{pseudolinearpositivity}. This begs the question: \\

\textit{Can consistent UV physics ever produce (even in principle) a massive gravitino with potential \eqref{quartics}?}\\
 
To answer this question, we will study the on-shell 2-to-2 scattering amplitudes between these massive gravitinos. \\

\noindent {\bf Positivity bounds -}
Dispersion relation arguments can be used to relate the above assumptions (of consistent physics in the UV) to properties that the EFT Wilson coefficients inherit in the IR: so-called \emph{positivity bounds} \cite{Adams:2006sv} (see also \cite{MassiveGravityPositivity2, Jenkins:2006ia,  Adams:2008hp, Nicolis:2009qm, Bellazzini:2014waa, Bellazzini:2015cra, Baumann:2015nta, Bellazzini:2016xrt, Cheung:2016yqr,Bonifacio:2016wcb,deRham:2017avq,deRham:2017imi,deRham:2017zjm,Bellazzini:2017fep,deRham:2017xox,deRham:2018qqo, Bellazzini:2019xts} for recent developments). If these bounds are violated, it means that there is no way to add new heavy degrees of freedom in the UV to repair the apparent violations of unitarity/causality in the IR. 
The simplest such positivity bound is,
\begin{equation}
 \partial_s^2 \mathcal{A} (s, t) \big|_{t=0} \geq 0 
 \label{eqn:pos}
 \end{equation}
for any 2-to-2 elastic scattering amplitude $\mathcal{A}$ (as a function of Mandelstam invariants $s$ and $t$), computed within the EFT and with low energy poles and branch cuts removed. ``Elastic'' refers to the outgoing particles sharing the same quantum numbers as the ingoing particles (i.e. same species, same chirality, same helicity, etc.). 
We provide a brief derivation of this bound, and describe its recent extensions, in Appendix B. 
Ultimately, if one can find a scattering process and an energy $s$ (within the EFT) for which \eqref{eqn:pos} is violated, then this rules out any possibility of UV completing the EFT with standard heavy physics.   \\

\noindent {\bf Gravitino -}
The polarisation tensors for a massive spin-3/2 field are constructed in Appendix B. Ingoing left-handed particles are denoted $X_{\alpha \beta \dot{\alpha}}^{(h)}$ and ingoing right-handed particles are denoted $\bar{Y}_{\alpha \dot{\alpha} \dot{\beta} }^{(h)}$, where $h$ is the helicity.
At high energies, it is $X^{(-1/2)}$ and $\bar{Y}^{(+1/2)}$ that grow like $s^{3/4}$ (and behave like the components of a Goldstone mode, $d \lambda$ and $d \bar{\lambda}$). 
The only elastic helicity amplitudes, $\mathcal{A}^{AB\to AB}_{h_A h_B h_A h_B} (s,t)$, from the potential \eqref{quartics} growing as fast as $s^2$ at high energy (and thus constrained by positivity) are
\begin{widetext}
\begin{align}
\mathcal{A}^{X X \to X X}_{-1/2, -1/2, -1/2,-1/2}   &=  \frac{  2g_7 + g_8 }{ m_{3/2}^3 \mathcal{M}_{3/2}^3 } \left[    
- \frac{2}{9}  s^3 +  \frac{4}{3} m_{3/2}^2 s^2 + ...    
\right] -  \frac{  2 g_8 }{ m_{3/2}^3 \mathcal{M}_{3/2}^3 } \left[  \frac{2}{9} s^2 t   + ... \right]   \label{XXXXhel} \\
 \mathcal{A}^{X Y \to X Y}_{-1/2, +1/2, -1/2, +1/2}  &= \frac{  2g_7 + g_8 }{ m_{3/2}^3 \mathcal{M}_{3/2}^3 } \left[    
- \frac{2}{9}  s^3 +  \frac{4}{3} m_{3/2}^2 s^2 + ...    
\right]  +  \frac{  6 g_7 + g_8 }{ m_{3/2}^3 \mathcal{M}_{3/2}^3 } \left[  \frac{2}{9} s^2 t  + ...  \right]. \label{XYXYhel}
\end{align}
\end{widetext}
Since on-shell polarisations are traceless\footnote{For a pedagogical introduction to the on-shell constraints of massive spinning particles we refer the reader to \cite{RahmanHigherSpin}.}, it is only $g_7$ and $g_8$ which give non-vanishing contributions to elastic 2-to-2 scattering. 
The simplest positivity bound
\eqref{eqn:pos}
leads to the constraint,
\begin{equation}
 g_8 = - 2 g_7 \; \, .
 \label{GoldstonePositivity}
\end{equation}
In this region of parameter space, the decoupling limit interactions $(\partial \lambda)^4$ are consistent with positivity\footnote{
Once $g_8 = -2g_7$, the forward limit scattering amplitude \emph{vanishes} at this order, which for exactly massless particles signals an issue since then $\partial_s^2 \mathcal{A} (s,0)$ cannot be \emph{strictly} positive, see also \cite{Bellazzini:2016xrt}. In our case, however, $\lambda$ is not exactly massless, and the positivity bound \eqref{eqn:pos} is not a strict inequality since we have neglected subleading $\mathcal{O} (m_{3/2}/\mathcal{M}_{3/2}$) corrections (which may have the correct sign to satisfy the bound).   
}. However, these interactions are not generated in isolation: they are accompanied by a mixing between $\lambda$ and the transverse modes which grows like $\sim s^2$ and must also obey positivity. For helicity eigenstates, selection rules set to zero many such amplitudes, but for superpositions of helicity eigenstates positivity of this mixing demands,
\begin{equation}
g_7 = g_8 = 0 \, ,
\label{eqn:nog7g8}
\end{equation}
i.e. there can be no UV completion of these operators (the explicit amplitudes are deferred to Appendix B).  
Standard heavy physics in the UV can never give rise to a single massive gravitino with quartic self-interactions in the IR (at least not those which contribution to elastic 2-to-2 scattering at tree level and obey an NDA power counting scheme \eqref{eqn:Vcounting}).  \\

\noindent
{\bf Tensor -} 
Let us compare with the positivity situation for massive spin-2.
Since the on-shell spin-2 polarisations are also traceless, from the quartic potential \eqref{eqn:spin2quartics} only $d_1$ and $d_3$ contribute to 2-to-2 scattering. 
From the scattering of four Goldstone (helicity-0) modes, 
\begin{align}
\partial_s^2  \mathcal{A}_{0000} (s, t) \big|_{t=0} = \frac{16}{3 \mathcal{M}_2^2 m_2^6 } \Bigg[&  (d_1 + 2 d_3 ) s (s-4m_2^2)  \nonumber \\
&+ 4 ( 4 d_1 + 7 d_3 ) m_2^4   \Bigg] 
\end{align}  
positivity \eqref{eqn:pos} for a wide range of $s$ requires the tuning,
\begin{equation}
 d_1 = - 2 d_3 \, .
 \label{eqn:spin2postuning}
\end{equation}
c.f. \eqref{GoldstonePositivity}.
This coincides with (the on-shell part of) the tuning in \eqref{eqn:hhhh} which raises the cutoff. The same conclusion also applies to the cubic coefficients $c_1$ and $c_2$, the existence of a standard UV completion enforces the tuning which raises the cutoff. 
Going beyond the purely longitudinal sector, one finds that with the tuning \eqref{eqn:spin2postuning}, all of the definite helicity amplitudes can satisfy positivity if $d_3 < 0$ \cite{pseudolinearpositivity}\footnote{
In a massive theory, the zero on the right hand side of the positivity bounds should always be read as $\mathcal{O} ( m_S / \mathcal{M}_S)$, since we have discarded higher derivative interactions which will give (possibly positive) subleading contributions to the amplitude, and so with that in mind having an amplitude saturate the bounds is not an issue.  
}. However once superpositions of different helicity states are considered (e.g. the transversity states described in Appendix B), then positivity requires,
\begin{equation}
 d_1 = d_3 = 0 
 \label{eqn:spin2d1d3}
\end{equation}
and there is no allowed region of parameter space (c.f. \eqref{eqn:nog7g8}), at least for a single massive spin-2 EFT whose Abelian gauge symmetry is broken in the NDA power counting \eqref{eqn:Vcounting}. 
%

Interestingly, the same is not true for the non-Abelian symmetry (i.e. dRGT massive gravity).
If one momentarily relaxes the Abelian NDA counting, and allows for two-derivative interactions to compete with the zero-derivative potential, then there is at least one tuning which \emph{does} satisfy positivity: namely the quartic truncation of Einstein-Hilbert plus the dRGT potential terms (whose amplitudes were described in \cite{Aubert:2003je} and \cite{MassiveGravityPositivity1} respectively). 
In fact, once the two-derivative terms have reorganised into an EFT of the form \eqref{eqn:Vcounting}, but with non-Abelian diffeomorphisms controlling the power counting, it was shown in \cite{MassiveGravityPositivity2} that positivity requires the special tuning which raises the strong coupling to $\Lambda_3$, just like \eqref{eqn:spin2postuning} in the Abelian case. However, the difference is that the cubic and quartic two-derivative terms from the Einstein-Hilbert kinetic term (now enhanced relative to the Abelian NDA counting because they come from $\mathcal{L}_{\rm gauge}$ rather than $V$) are sufficiently positive that they replace \eqref{eqn:spin2d1d3} with an island of allowed values \cite{MassiveGravityPositivity1, MassiveGravityPositivity2} (which depends on the weak coupling that controls EFT loops \cite{deRham:2017xox, Bellazzini:2017fep}).  

In contrast, for a single spin-3/2 field it would not be possible for two-derivative terms to compete with zero-derivative terms without permanently violating the NDA power counting \eqref{eqn:Vcounting}, since there is no non-Abelian gauge symmetry to reorganise the large derivative couplings (this is because any anti-commutator between fermionic generators must give back a bosonic generator and so a gauged non-Abelian fermionic symmetry requires the presence of a bosonic gauge field).
\\

\noindent {\bf Vector -}
Consider how the severe positivity constraints \eqref{eqn:nog7g8} and \eqref{eqn:spin2d1d3} compare with the spin-$1$ case. 
For a massive vector, with potential \eqref{Vspin1}, the high-energy amplitude between four longitudinal polarisations is simply, 
\begin{equation}
\partial_s^2 \mathcal{A}_{0000} = \frac{16}{ \mathcal{M}_1^2 m_1^2} \, C_1
\end{equation}
and so the analogous positivity constraint on the Goldstone sector yields $C_1 > 0$.
For the massive tensor and gravitino, it was the $s$ dependence of $\partial_s^2 \mathcal{A}$ which turned the bound \eqref{eqn:pos} into an equality, but in the spin-$1$ case the fastest growing amplitude is only $\sim s^2$ at leading order (so any positive $C_1$ satisfies \eqref{eqn:pos} for all $s$). 
Crucially, the amplitudes which mix longitudinal and transverse modes only grow $\sim s$ and are therefore not constrained by positivity. Once the Goldstone sector satisfies positivity, no further tuning is necessary. 
This is perfectly consistent with known UV completions like the Higgs mechanism, in which all that is required for a healthy massive vector in the IR is for a single scalar Goldstone to be eaten by a massless vector. 
\\

Summing up, we have found that the massive gravitino fits neatly into place between the vector and tensor: like the massive vector, there is no way to raise the strong coupling scale; and like the massive graviton, positivity of the mixing between longitudinal and transverse modes forces any interactions which break an Abelian gauge symmetry to be set to zero. However, unlike for the spin-$2$ case, there is no non-Abelian gauge symmetry (with the required additional interactions) to save the day. 
%
Ultimately, it is not possible to generate natural (order one) self-interactions for a massive spin-3/2 particle in the tree-level 2-to-2 amplitude from UV physics which is unitary, analytic, local and Lorentz invariant.

\section{Conclusions \& outlook}
\noindent The goal of this paper has been to elucidate the UV constraints on the IR physics of isolated massive particles of specific spin, with emphasis on the novel spin-$3/2$ case. Our results can also be read in reverse, as the UV implications of a detection of an isolated massive and spinning mode in the IR.

\begin{table*}
\begin{tabular}{l  l | l l | l l }
Spin & LO operators  & $\Lambda$ & Raise $\Lambda$?  & Positivity: Goldstone  & Positivity: Mixing  \\
\hline
1  &  $C_1$  &  $( m \mathcal{M} )^{1/2}$  &  $\xmark$    					& $\cmark \,( C_1 >0)$				& N/A since amplitudes $\sim s$   \\
3/2  & $g_{1-8}$  &  $ ( m \mathcal{M} )^{1/2}$ &  $\xmark$ 					& 	$\cmark \, (g_8=-2g_7)$	&  $\xmark \, (g_7=g_8=0)$     \\
$2_{A}$ &   $ (c_{1-3}), \, d_{1-5} $  & $( m^4 \mathcal{M} )^{1/5}$  & $\cmark$	& $\cmark \, (d_1=-2d_3)$					& $\xmark \, (d_1 = d_3 = 0) $    \\
$2^*_{A}$ &  $(c_3), d_5$   & $( m^2 \mathcal{M} )^{1/3}$  &  $\xmark$			& 	$\cmark $					& $ \xmark $ \cite{pseudolinearpositivity}  \\
$2_{NA}$ &   $(\tilde{c}_{1-3}), \, \tilde{d}_{1-5} $  & $( m^4 \mathcal{M} )^{1/5}$  & $\cmark$	& $\cmark$					& $\xmark$ (unless tuned to $*$ \cite{MassiveGravityPositivity2})     \\
$2^*_{NA}$ &  $(\tilde{c}_3), \tilde{d}_5$   & $( m^2 \mathcal{M} )^{1/3}$  &  $\xmark$			& 	$\cmark $					& $ \cmark $\, 
\cite{MassiveGravityPositivity1} \\

\end{tabular}
\caption{
UV implications of massive spinning particles, for spins $1$, $3/2$ and $2$. 
For spin-$2$ there are two cases, depending on whether it is an Abelian ($2_{A}$) or non-Abelian ($2_{NA}$) gauge symmetry which is softly broken by the mass terms. 
The leading order quartic (and cubic) unitarity gauge potential, $m^{2-\theta} \mathcal{M}^{2+\theta} \, V$, contains the Wilson coefficients given in the second column. 
The strong coupling scale can be established off-shell at the level of the Lagrangian, and only for spin-$2$ can it be raised by an $\mathcal{O} (1)$ tuning of the Wilson coefficients (in terms of, e.g., $c_3$ and $d_5$), leading to theories with a particular WZ structure ($2^*$). 
In all cases, the purely longitudinal (Goldstone) sector can satisfy positivity bounds for some region of parameter space, but spin-$1$ is the only Abelian theory whose mixing between longitudinal and transverse modes is also consistent with positivity (we implicitly refer only to those operators that contribute to the on-shell 2-to-2 scattering amplitude at tree-level). For the non-Abelian spin-2, it is the particular two-derivative terms (comparable to the zero-derivative potential) which allow for all known positivity bounds to be satisfied. 
}
\end{table*}

With regards to positivity bounds we find a similarity between the graviton and the gravitino; in both cases, there are severe constraints on the possible IR coefficients from positivity bounds involving either solely the longitudinal mode, or the longitudinal and transverse modes (arising from the kinetic demixing). These constraints rule out generic interactions for a spin-2 field (i.e.~such interactions do not have a UV completion that is unitary, etc.); however, there is a specific tuning of the coefficients that evades these constraints and presents a viable IR description, as long as the kinetic sector is of the Einstein-Hilbert form rather than its linearisation. For the gravitino, we show that the positivity bounds are equally severe and that there is no specific tuning that evades them. In contrast, for a vector the positivity bounds are weaker due to the absence of kinetic demixing, thus allowing for a consistent theory even without the possibility to tune the interactions. Moreover, this is the only case with a known UV completion in the form of the Higgs mechanism. The different constraints are summarised in Table 1.

Our strong coupling analysis of Section II applies off-shell, and so although we have considered only quartic operators it represents a necessary condition for raising the strong coupling scale for any $n$-point amplitude. 
There we focussed on the longitudinal sector, which becomes an effective shift-symmetric Goldstone theory at high energies, dominated by purely longitudinal interactions. Interestingly here it is the vector and gravitino that work in a similar way with one unable to raise the strong coupling scale by tuning the Wilson coefficients, in contrast to the massive graviton. As we have shown, the crucial difference is that only the massive graviton's longitudinal mode has Wess-Zumino structures for its self-interactions. 

We have worked within the simplest EFT power counting \eqref{eqn:Vcounting}, which amounts to a particular assumption about how the UV fields give rise to the massive spinning particle in the IR. 
This is not the only radiatively stable power counting scheme available: for instance a ``single-scale--single-coupling'' scheme \cite{Georgi:1989xy, Giudice:2007fh}, in which higher spin fields always appear as $g_* \Phi$ with a weak coupling $g_* \sim (m_S / \mathcal{M}_S )^S$ (which ensures that all \Stuckelberg fields appear with $\partial / \mathcal{M}_S$ rather than $\partial / m$) could be used to ensure that perturbative unitarity is not violated until $\mathcal{M}_S$, as advocated in \cite{MassiveGravity2}. However, in any power counting scheme in which the zero-derivative potential dominates scattering processes, one must contend with strict positivity requirements. Our results suggest that simple NDA (or NDA-like) power counting schemes are not suitable for describing massive higher spin fields (with weakly broken Abelian symmetries) in the IR with consistent UV completions.   
Attempting to artificially suppress the zero-derivative interactions (e.g. by tuning all of their Wilson coefficients by hand) often leads to tension with more refined positivity bounds which subtract loop-contributions within the EFT \cite{deRham:2017imi,Bellazzini:2017fep}.

Our analysis implies the simple yet strong result that {\it a massive spin-$3/2$ field cannot exist in isolation}. Remarkably, a similar conclusion was reached for the massless case, which was shown to necessarily belong to a supermultiplet \cite{SoftGravitino}. A natural interpretation of our result implies a similar structure in the massive case. There could be a massive supergravity multiplet where the quartic potentials for the massive graviton and gravitino could form a supersymmetry invariant, and their longitudinal modes could be the dim-10 super-Galileon structures of e.g.~\cite{Farakos, Elvang}. Alternatively, supersymmetry could be spontaneously broken, with a massless graviton and a massive gravitino that acquires a mass at the SUSY breaking scale via the super-Higgs mechanism. We leave such questions for future investigations.

An obvious extension of our work would be to consider higher spinning fermionic particles. Higher spinning bosons were considered in \cite{Bonifaciothesis,BrandoHS,BonifacioHinterbichler1}. We expect that our approach of classifying possible off-shell currents in order to construct total derivatives and WZ terms (and hence specific interactions with a raised strong coupling scale, as was crucial for the spin-2 case) will prove equally valuable in this context. An obvious way to raise the strong coupling scale for a massive fermion with spin $\geq 5/2$ would be to trivially add spinor indices to the potentials derived in \cite{Bonifaciothesis,BrandoHS}, but a study of WZ interactions may lead to a more richer structure requiring Pauli matrices.

Interestingly, it was shown in \cite{BrandoHS} that a massive spin-$3$ particle with a broken Abelian gauge symmetry violates positivity bounds due to mixings between transverse and longitudinal modes, as for (Abelian) spin-$2$ and spin-$3/2$. It would be very interesting if this was a generic feature i.e. if positivity bounds in the presence of mixings between transverse and longitudinal modes can only be satisfied for spin $\geq 3/2$ if the kinetic sector enjoys a non-Abelian gauge symmetry. As in the massless case, this would make spin-$2$ special. We leave this for future work.


\section*{Acknowledgments}
\noindent We would like to thank Sadra Jazayeri, Silvia Nagy, Antonio Padilla, Enrico Pajer, Massimo Porrati, Rakibur Rahman and Pelle Werkman for useful discussions. SM is supported by an Emmanuel College Research Fellowship, and also in part by STFC consolidated grant ST/P000681/1. DR acknowledges the Dutch funding agency ‘Netherlands Organisation for Scientific Research’ (NWO) for financial support. DS acknowledges the research program VIDI with Project No. 680-47-535, which is (partly) financed by the Netherlands Organisation for Scientific Research (NWO), for financial support.

\appendix
\section{Appendix A: Off-shell details}

\noindent We can convert the massive gravitino potential into a more familiar basis with a mixture of $SO(1,3)$ and $SU(2) \times SU(2)$ indices by making use of the Pauli matrices. A complete basis requires only two ($\eta$-traceful) tensor bi-linears, one without any Pauli matrices and one using the anti-symmetric combination $4(\sigma^{\mu\nu})_{\alpha}{}^{\beta} = (\sigma^{\mu})_{\alpha \dot{\alpha}} (\bar{\sigma}^{\nu})^{\dot{\alpha}\beta} - (\sigma^{\nu})_{\alpha \dot{\alpha}} (\bar{\sigma}^{\mu})^{\dot{\alpha}\beta}$. The tensor bi-linears are $\psi_{\mu} \psi_{\nu}, \psi_{(\mu} \sigma_{\nu) \rho} \psi^{\rho}$, with equivalent structures for $\bar{\psi}\bar{\psi}$. Out of these two tensors we can construct four Lorentz irreps by taking the trace and traceless parts. We have\footnote{
Note that the gravitino mass term is precisely $S^{(2)}$ since this becomes a total derivative for $\psi_{\mu} = \partial_{\mu} \lambda$.} 
\begin{align}
S^{(0)} &= \psi^{\mu}\psi_{\mu}, \\ 
S^{(2)} &= \psi^{\mu}\sigma_{\mu\nu}\psi^{\nu}, \\
T^{(0)}_{\mu\nu} &= \psi_{\mu}\psi_{\nu} - \frac{1}{4}\eta_{\mu\nu}S^{(0)}, \\
T^{(2)}_{\mu\nu} &= \psi_{(\mu} \sigma_{\nu) \rho} \psi^{\rho} - \frac{1}{2}\eta_{\mu\nu} S^{(2)}.
\end{align} 
The eight quartic interactions can now be written as
\begin{align}\label{quartics2}
\mathcal{L}_{(4)} &= (s_{00} S^{(0)}S^{(0)} + s_{02}S^{(0)}S^{(2)} + c.c) \nonumber \\ &+ \hat{s}_{00}S^{(0)}\bar{S}^{(0)} 
 + (\hat{s}_{02}S^{(0)}\bar{S}^{(2)} + c.c) \nonumber \\ &+ \hat{s}_{22} S^{(2)}\bar{S}^{(2)}
+ t_{00}T^{(0)} \cdot \bar{T}^{(0)} \nonumber \\ &+ (t_{02} T^{(0)} \cdot \bar{T}^{(2)} + c.c) + t_{22}T^{(2)} \cdot \bar{T}^{(2)}.
\end{align}
In principle we could have also included a $S^{(2)}S^{(2)}$ term but by a Fierz identity this term is a linear sum of the first two in \eqref{quartics2}. We could have also made use of the bi-linear $\psi_{[\mu} \sigma_{\nu]\rho}\psi^{\nu}$ but again any interaction constructed from this bi-linear and its complex conjugate is already containined in \eqref{quartics} thanks to Fierz identities. 

We now show that when $\psi_{\mu \alpha} = \partial_{\mu} \lambda_{\alpha}$, the $\lambda$ equation of motion is only trivial if each coefficient in the unitary gauge potential vanishes i.e we show that there is no non-trivial dim-$10$ total derivative. First consider the operators that break chiral symmetry i.e. with $s_{00}$ and $s_{02}$ couplings. There is clearly no way these two terms could combine into a total derivative since terms with zero or two sigma matrices decouple and the two terms are themselves non-trivial. We must therefore fix $s_{00} = s_{02} = 0$ if these contributions to the equation of motion are to be made trivial.

Moving onto the remaining six terms, the $\lambda^{\alpha}$ equation of motion contains schematically two different types of terms: one containing $\partial \partial \lambda$ the other containing $\partial \partial \bar{\lambda}$. Within these two sets, these derivatives can form a traceless or pure trace combination. Initially consider the traceless $\partial \partial \lambda$ terms. Clearly the scalar bi-linears cannot contribute, and neither can the $\bar{t}_{02}$ and $t_{22}$ terms due to the symmetry/Grassman properties of $T^{(2)}_{\mu\nu}$. It is then clear that the only contributions, $t_{00}$ and $t_{02}$, cannot cancel since in each case the traceless $\partial \partial \lambda$ is contracted with either $\bar{T}^{(0)}_{\mu\nu}$ or $\bar{T}^{(2)}_{\mu\nu}$ which are two linearly independent tensors. We therefore need to fix $t_{00} = t_{02} = 0$. Now consider the pure trace $\Box \lambda$ terms where only $\hat{s}_{00}$ and $\hat{s}_{02}$ contribute. The $\bar{\hat{s}}_{02}$ and $\hat{s}_{22}$ terms do not contribute due to the symmetry properties of $S^{(2)}$. It follows that we must fix $\hat{s}_{00} = \hat{s}_{02} = 0$ since the two $\Box \lambda$ contributions are multiplied by the linearly independent scalars $\bar{S}^{(0)}$ and $\bar{S}^{(2)}$. 

Finally, consider the $\partial \partial \bar{\lambda}$ terms. We only have the $\hat{s}_{22}$ and $t_{22}$ terms remaining and the total contribution to the equation of motion is
\begin{align}
&\partial_{\mu}\left(\frac{\partial \mathcal{L}^{(4)}}{\partial \partial_{\mu} \lambda^{\alpha}}\right) = -2 \hat{s}_{22} (\sigma_{\mu\nu})_{\alpha}{}^{\beta} \partial^{\nu}\lambda_{\beta} \partial^{\mu}\bar{S}^{(2)} \nonumber \\
&- t_{22} (\sigma_{\nu \rho})_{\alpha}{}^{\beta} \partial^{\rho}\lambda_{\beta} \partial_{\mu} \bar{T}^{(2) \mu\nu}+ t_{22} (\sigma_{\nu \rho})_{\alpha}{}^{\beta} \partial_{\mu}\lambda_{\beta} \partial^{\rho} \bar{T}^{(2) \mu\nu}.
\end{align}
Clearly $\partial_{\mu}\bar{S}^{(2)}$ and $\partial_{\mu}T^{(2)\mu\nu}$ live in the vector representation of the Lorentz group, whereas $\partial^{\rho}T^{(2)\mu\nu}$ contains a piece transforming in the traceless hook representation. This is non-zero and cannot be cancelled so we are required to set $t_{22} = 0$ and therefore $\hat{s}_{22} = 0$. 

We therefore conclude that there are no combinations of dim-$10$ operators that combine into a total derivative and therefore no way to raise the strong coupling scale for a massive gravitino without making the theory free.


\section{Appendix B: On-shell details}

\noindent In this Appendix, we provide details of the on-shell 2-to-2 scattering amplitudes for spin-3/2 particles, and the positivity bounds which they must satisfy if a ``standard\footnote{
By ``standard'', we mean unitary, analytic, Lorentz-invariant and obeys the Froissart bound.
}'' UV completion is to exist.   
\\

\noindent
{\bf Polarisations - }
With the conventions of \cite{Conventions}, the on-shell polarisations for an incoming spin-1/2 particle with mass $m$, momentum $\mathbf{p} = k ( \sin \theta , 0, \cos \theta )$ and helicity $h$ are,
\begin{align}
& \text{LH, $h=+1/2$:}  &&x_{\alpha}^{+} = \frac{m + \omega_k - k}{\sqrt{2} \sqrt{ m + \omega_k} } \left( \begin{array}{c}
\cos (\theta/2)  \\
\sin (\theta/2 ) 
\end{array} \right)  \;\; ,   \\
& \text{LH, $h=-1/2$:}  &&x_{\alpha}^{-} =  \frac{m + \omega_k + k}{\sqrt{2} \sqrt{ m + \omega_k} } \left( \begin{array}{c}
- \sin ( \theta/2)  \\
\cos (\theta/2)
\end{array} \right) \;\; ,  
\end{align}
\begin{align}
& \text{RH, $h=+1/2$:} &&\bar{y}_{\dot \alpha}^+ = \frac{m + \omega_k + k}{\sqrt{2} \sqrt{ m + \omega_k} } \left( \begin{array}{c}
 -  \sin ( \theta / 2 )  \\
 \cos ( \theta / 2 )
\end{array} \right) \;\; ,    \\
& \text{RH, $h=-1/2$:} &&\bar{y}_{\dot \alpha}^- =  \frac{m + \omega_k - k}{\sqrt{2} \sqrt{ m + \omega_k} } \left( \begin{array}{c}
- \cos ( \theta/2)  \\
\sin (\theta/ 2)
\end{array} \right) \;\; ,    
\end{align}
where $\omega_k = \sqrt{k^2 + m^2}$. The analogous outgoing states\footnote{
There is a CPT relation between an outgoing RH helicity $h$ fermion and an incoming LH helicity $-h$ fermion, namely $y^h_{\alpha} = i (-1)^{-h} x^{-h}_{\alpha}$, but we shall keep all four symbols distinct for clarity. 
} are $\bar{x}_{\dot \alpha}^{\pm}$ and $y_{\alpha}^{\pm}$, and the field is canonically quantized\footnote{
Note that it is the operators $\hat{a}_{\mathbf{p},h}, \, \hat{a}_{\mathbf{p},h}^{\dagger}$ which anti-commute, and so $x^h_{\alpha}$ and $y^h_{\alpha}$ may be regarded as commuting spinors
} as $\lambda_{\alpha} (x) = \int_p \sum_h \,\left(  x^{\mathbf{p} , h}_\alpha \hat{a}_{\mathbf{p},h} e^{-i p \cdot x} + y^{\mathbf{p} ,h}_{\alpha} \hat{a}_{\mathbf{p},h}^{\dagger}   e^{i p \cdot x }  \right)$, where the superscripts in $x^{\mathbf{p},h}_\alpha$ denote the momentum and helicity used to construct the polarisation, and $\int_p$ is the usual integral over on-shell, future-pointing $p^\mu$. 
 
Any higher spin polarisation tensor can be constructed by applying raising/lowering operators to the highest weight state. For instance, for an ingoing LH spin-3/2 particle the different helicity states are, 
\begin{align}
m X_{\alpha \beta \dot \alpha}^{+3/2} &=   x^{+}_{\alpha}  x^{+}_{\beta} \bar{y}^{+}_{\dot \alpha} \nonumber \\
m X_{\alpha \beta \dot \alpha}^{+1/2} &= \frac{1}{\sqrt{3}} \left( 
 x^{-}_{\alpha}  x^{+}_{\beta} \bar{y}^{+}_{\dot \alpha} 
+  x^{+}_{\alpha}  x^{-}_{\beta} \bar{y}^{+}_{\dot \alpha}
 +  x^{+}_{\alpha}  x^{+}_{\beta} \bar{y}^{-}_{\dot \alpha}   \right)  \nonumber \\
 m X_{\alpha \beta \dot \alpha}^{-1/2} &= \frac{1}{\sqrt{3}} \left( 
 x^{+}_{\alpha}  x^{-}_{\beta} \bar{y}^{-}_{\dot \alpha} 
+  x^{-}_{\alpha}  x^{+}_{\beta} \bar{y}^{-}_{\dot \alpha}
 +  x^{-}_{\alpha}  x^{-}_{\beta} \bar{y}^{+}_{\dot \alpha}   \right)  \nonumber \\
m X_{\alpha \beta \dot \alpha}^{-3/2} &= x^{-}_{\alpha}  x^{-}_{\beta} \bar{y}^{-}_{\dot \alpha}  
\label{Xpols}
\end{align}
and similarly for an ingoing RH spin-3/2 particle,
\begin{align}
m \bar{Y}_{\alpha \dot \beta \dot \alpha}^{+3/2} &=   x^{+}_{\alpha}  \bar{y}^{+}_{\dot \beta} \bar{y}^{+}_{\dot \alpha}  \nonumber \\
m \bar{Y}_{\alpha \dot \beta \dot \alpha}^{+1/2} &= \frac{1}{\sqrt{3}} \left( 
 x^{-}_{\alpha}  \bar{y}^{+}_{\dot \beta} \bar{y}^{+}_{\dot \alpha} 
+  x^{+}_{\alpha}  \bar{y}^{-}_{\dot \beta} \bar{y}^{+}_{\dot \alpha}
 +  x^{+}_{\alpha}  \bar{y}^{+}_{\dot \beta} \bar{y}^{-}_{\dot \alpha}   \right) \nonumber \\
m \bar{Y}_{\alpha \dot \beta \dot \alpha}^{-1/2} &= \frac{1}{\sqrt{3}} \left( 
 x^{+}_{\alpha}  \bar{y}^{-}_{\dot \beta} \bar{y}^{-}_{\dot \alpha} 
+  x^{-}_{\alpha}  \bar{y}^{+}_{\dot \beta} \bar{y}^{-}_{\dot \alpha}
 +  x^{-}_{\alpha}  \bar{y}^{-}_{\dot \beta} \bar{y}^{+}_{\dot \alpha}   \right)  \nonumber \\
m \bar{Y}_{\alpha \dot \beta \dot \alpha}^{-3/2} &=  x^{-}_{\alpha}  \bar{y}^{-}_{\dot \beta} \bar{y}^{-}_{\dot \alpha}  \, , 
\label{Ypols}
\end{align}
where again the corresponding outgoing states\footnote{
We have chosen a phase convention in which the spin-3/2 states satisfy the analogous CPT relation, $Y^{h}_{\alpha \beta \dot \alpha} = i (-1)^{-h} X^{-h}_{\alpha \beta \dot \alpha}$.
} are simply $\bar{X}_{\dot \alpha \dot \beta \alpha}$ and $Y_{\dot \alpha \beta \alpha}$, and we canonically quantise the field as $\Psi_{\alpha \beta \dot \alpha} = \int_p \sum_h \,\left(  X^{\mathbf{p} , h}_{\alpha \beta \dot \alpha} \hat{a}_{\mathbf{p},h} e^{-i p \cdot x} + Y^{\mathbf{p} ,h}_{\dot \alpha \beta \alpha} \hat{a}_{\mathbf{p},h}^{\dagger}   e^{i p \cdot x }  \right)$. 
\\

\noindent
{\bf Helicity amplitudes - }
We compute the amplitude for $AB \to CD$ scattering processes in terms of the Mandelstam invariants\footnote{
This is most easily done by first going to the center-of-momentum frame, where each $\mathbf{p}_A = k ( \sin \theta_A , 0 , \cos \theta_A )$ and $\theta_1 = 0 \, , \; \theta_2 = \pi \, , \; \theta_3 = \theta \, , \; \theta_4 = \pi + \theta$. The $k$ and $\theta$ are then related to the Mandelstam variables by, $k = \sqrt{ s - 4m^2}$, $\cos \theta = 1 + 2 t / (s-4m^2)$. Once this replacement has been performed, the resulting function of the invariants $s$ and $t$ is frame-independent.  
} $s = -(p_A + p_B)^2$, $t= - (p_A+p_C)^2$ and $u = - (p_A+p_D)^2$, which obey the relation $s+t+u=4m^2$ when the particles are on-shell.  
We first consider scattering the helicity states \eqref{Xpols} and \eqref{Ypols}, and denote the resulting amplitude $\mathcal{A}^{A B \to CD}_{h_A h_B h_C h_D} (s,t,u)$.
 
Scattering states of definite helicity has a number of advantages. Firstly, the polarisations have a convenient high-energy scaling with $s$ (at fixed $t$),
\begin{align}
& X^{3/2}_{\alpha \beta \dot \alpha} \sim s^{-\frac{1}{4} }  , \;  X^{1/2}_{\alpha \beta \dot \alpha} \sim s^{\frac{1}{4}}  ,   X^{-1/2}_{\alpha \beta \dot \alpha} \sim s^{\frac{3}{4}}  , \;  X^{-3/2}_{\alpha \beta \dot \alpha} \sim s^{\frac{1}{4}}  \nonumber \\
& \bar{Y}^{3/2}_{\alpha \beta \dot \alpha} \sim s^{\frac{1}{4}}  , \;  \bar{Y}^{1/2}_{\alpha \beta \dot \alpha} \sim s^{\frac{3}{4}} \; , \;\;\;  \bar{Y}^{-1/2}_{\alpha \beta \dot \alpha} \sim s^{\frac{1}{4}}  , \;  \bar{Y}^{-3/2}_{\alpha \beta \dot \alpha} \sim s^{-\frac{1}{4}}  \label{XYgrowth}
\end{align}
allowing one to immediately conclude that it is the processes like $X^{-1/2} X^{-1/2} \to X^{-1/2} X^{-1/2}$ (whose tree-level amplitude $\mathcal{A} \sim s^3$) which present the most severe violation of perturbative unitarity. Second, invariance under $C$, $P$ and $T$ gives rise to very simple selection rules on helicity amplitudes---for instance, while $X^{-1/2} X^{+1/2} \to X^{-1/2} X^{+1/2}$ may seem to grow like $\sim s^2$ from \eqref{XYgrowth}, it actually vanishes identically. In the helicity basis, the only elastic amplitudes\footnote{
For elastic processes like $X X \to X X$ (i.e. two incoming LH spin-3/2 particles scattering into two outgoing LH spin-3/2 particles), the only vertices from \eqref{quartics} which contribute are $g_7$ and $g_8$ (since $g_1, g_2$ do not contain the necessary $X X \bar X \bar X$ terms and $g_{3-6}$ contain traces of $\Psi$, which vanish on-shell since physical polarisations are always traceless, c.f. \eqref{Xpols} and \eqref{Ypols}).  
} which grow like $s^2$ or faster are $\mathcal{A}^{XX \to XX}_{-1/2,-1/2,-1/2,-1/2}$ and $\mathcal{A}^{XY \to XY}_{-1/2,+1/2,-1/2,+1/2}$, given in \eqref{XXXXhel} and \eqref{XYXYhel}.

However, the disadvantage of the helicity basis is that crossing symmetry is rather complicated at finite $t$ (crucially, the $u$-channel is not always a positive image of the $s$-channel, which makes positivity bounds difficult to construct). This was resolved in \cite{deRham:2017zjm} by instead using the transversity basis (i.e. scattering particular superpositions of helicity states) which enjoys a far simpler crossing relation. We will next describe these tranversity amplitudes, and then sketch how positivity arguments can be applied to constrain the EFT parameters $g_7$ and $g_8$.  
\\

\noindent
{\bf Transversity amplitudes - }
For a spin-3/2 particle, the transversity states are given by,
\begin{equation}
 X_{\alpha \beta \dot \alpha}^\tau = \sum_h u^{\tau}_h X^h_{\alpha \beta \dot \alpha} \;\;, \;\;  \bar{Y}_{\alpha \beta \dot \alpha}^\tau = \sum_h u^{\tau}_h \bar{Y}^h_{\alpha \beta \dot \alpha}
\end{equation}
where the spin-$3/2$ Wigner $u$ matrix is,
\begin{equation}
 u^h_{\tau} = 
\frac{1}{2\sqrt{2}} \left(
\begin{array}{cccc}
 1 & i \sqrt{3} & -\sqrt{3} & -i \\
 i \sqrt{3} & -1 & i & -\sqrt{3} \\
 -\sqrt{3} & i & -1 & i \sqrt{3} \\
 -i & -\sqrt{3} & i \sqrt{3} & 1 \\
\end{array}
\right) \, . 
\end{equation}
Scattering in this basis leads to the amplitudes, $\tilde{\mathcal{A}}_{\tau_A \tau_B \tau_C \tau_D}^{AB \to CD}$, given explicitly at the end of this Appendix in \eqref{Texplicit}. The advantage of this basis is that the $XY \to XY$ amplitudes obey the simple crossing relation,
\begin{equation}
\tilde{\mathcal{A}}^{XX \to XX}_{\tau_1 \tau_2 \tau_1 \tau_2} (s , t, u) = e^{i \chi (2 \tau_1 + 2 \tau_2 )} \tilde{ \mathcal{A} }^{XY \to XY}_{-\tau_1 -\tau_2 -\tau_1 -\tau_2} (u , t, s)
 \label{Tcrossing}
\end{equation}
with prefactor,
\begin{align}
 e^{i \chi} = \frac{ -s u + 2 i m \sqrt{s t u} }{ \sqrt{ s (s - 4 m^2) u (u - 4 m^2) } }  \, . 
\end{align}
This is much simpler than its helicity counterpart, and in particular maps a single $s$-channel amplitude to a single $u$-channel amplitude (with prefactor that can be made sign-definite). 
~\\

\noindent
{\bf Positivity bounds - }
The positivity bounds for elastic scalar amplitudes, $\mathcal{A} (s,t)$ are well-known \cite{Adams:2006sv}. In short, \emph{analyticity} allows one to express (derivatives of) the low energy amplitude in terms of a contour integral over the unknown UV amplitude,
\begin{align}
\partial_s^2 \mathcal{A} (s,t) &= \oint_s \frac{d \mu}{2 \pi i} \frac{\mathcal{A} (\mu,t) }{ (\mu - s )^3}  \nonumber \\
&\propto \int_{4m^2}^{\infty} \frac{d \mu}{\pi} \frac{\text{Im} \, \mathcal{A} (\mu,t)}{ (\mu - s )^3 } + \left( \mu \leftrightarrow 4m^2-\mu-t \right) \nonumber \\
&\quad + \int_{0}^{2 \pi} \frac{d \varphi }{2 \pi i} \frac{e^{-3 i \varphi} \mathcal{A} ( | \mu | e^{i \varphi} , t) }{ |\mu|^2 } \big|_{|\mu| \to \infty}
\label{contour}
\end{align}
The \emph{Froissart bound} (locality), $| \mathcal{A} ( s, t ) | < s \text{log}^2 s$ at large $|s|$, allows one to neglect the final line. By \emph{crossing symmetry} (Lorentz invariance), the $s$- and $u$-channel amplitudes are related by $\mathcal{A}_s (4m^2-s-t,t) = \mathcal{A}_u (s, t)$. 
Finally, since both remaining integrals are now proportional to $\text{Im} \mathcal{A} ( \mu , t)$ of some process, by \emph{unitarity} this must be positive in the foward limit (since $\text{Im} \mathcal{A} (\mu, 0) \geq | \mathcal{A} (\mu,0) |^2$ by the optical theorem\footnote{
Note that this is why the specialisation to elastic process only, i.e. processes of the form $AB \to AB$, is important---otherwise there is no optical theorem and $\text{Im} \, \mathcal{A}$ need not be positive.
}). Altogether, by specifying only the mild requirements of analyticity, unitarity, Froissart boundedness and Lorentz invariance in the full UV amplitude (but remaining agnostic as to what its field content, interactions, etc. are), we have established that\footnote{
The restriction to $0< s < 4m^2$ ensures that the denominators in \eqref{contour} are also positive.
},
\begin{equation}
 \partial_s^2 \mathcal{A} (s, 0 ) \geq 0 \;\;\;\; \text{ for all } \;\;\;\; 0 \leq s < 4m^2
\label{possimple}
\end{equation}
must be satisfied in the EFT at low energies.  
This has been generalised to particles with spin (in \cite{Adams:2008hp, Bellazzini:2014waa, Bellazzini:2016xrt}), away from the forward limit (in \cite{Nicolis:2009qm, deRham:2017avq} for scalars, and in \cite{deRham:2017zjm} for non-zero spin), and beyond the Mandelstam triangle by subtracting part of the branch cut from EFT loops (in \cite{deRham:2017xox, Bellazzini:2017fep}, which can also place new constraints on how weakly coupled the UV must be). 
 
Helicity amplitudes in the forward limit, $t=0$, exhibit the same analyticity and crossing properties as a scalar amplitude, and consequently \eqref{possimple} also applies. 
From the helicity amplitudes \eqref{XXXXhel} and \eqref{XYXYhel}, this gives,
\begin{equation}
 (2 g_7 + g_8 ) ( - s + 2 m^2  ) > 0 \, . 
\end{equation}
This is what provides the constraint \eqref{GoldstonePositivity} presented in the main text. There is only a single bound because the kinematic structure of $g_7$ and $g_8$ differs only at finite $t$.    
Strictly speaking, any departure from the forward limit introduces a complicated crossing relation for the helicity amplitudes which generally invalidates the above positivity argument, so it is not possible to use the $s^2 t$ pieces of the helicity amplitude to place further bounds on the remaining $g_7$. 
 
However, there are two possible extensions of \eqref{possimple} relevant for the present case. The first is to remain in the forward limit, and consider scattering \emph{arbitrary superpositions} of helicity eigenstates---this can be very constraining, but optimising the bounds over all possible superpositions is an NP-hard problem which is best implemented numerically. The second is to instead use transversity states (a particular superposition of helicities), whose crossing relation is sufficiently simple that positivity arguments can be applied at finite $t$.
In particular, the central result of \cite{deRham:2017zjm} is that the ``regulated transversity amplitude'', in this case given by,
\begin{align}
&\tilde{A}^+_{\tau_1 \tau_2 \tau_1 \tau_2} =  \nonumber \\
&s^3 (s - 4m^2 )^3 \left( \tilde{A}_{\tau_1 \tau_2 \tau_1 \tau_2} (s,t) + \tilde{A}_{-\tau_1 -\tau_2 -\tau_1 -\tau_2} (s, t)   \right) 
\end{align}
satisfies positivity bounds even beyond the forward limit, the simplest of which is\footnote{
The need to perform an additional six $s$ derivatives comes from the prefactor $s^3 ( s- 4m^2)^3$ we have used to remove unphysical poles from the transversity amplitudes. 
},
\begin{equation} \label{Tpositivity}
 f_{\tau_1 \tau_2} (s, t) = \frac{1}{8!} \partial_s^8 \left[  \tilde{A}^+_{\tau_1 \tau_2 \tau_1 \tau_2}      \right] \geq 0 
 \end{equation}
for all $s$ and $t$ in the Mandelstam triangle. 
There are no finite values of $g_7$ and $g_8$ which satisfy \eqref{Tpositivity}. 
Once $g_8 = - 2 g_7$ is imposed from the helicity bounds, it suffices to look at,
\begin{equation}
 f_{+3/2, +1/2} (s, 0) = +\frac{g_7}{3 m \mathcal{M}^{3}},\;\; f_{+3/2,-1/2} (s,0) = - \frac{g_7}{3m \mathcal{M}^{3}}
\end{equation}
to see that $g_7$ must be tuned to zero (at or least as small as the next subleading operator) in order to be consistent with positivity. This result could also have been obtained by scattering different arbitrary superpositions of helicities in the forward limit, but emerges immediately from the scattering transversity states.  \\

\noindent{\bf Explicit spin-3/2 amplitudes:}
Since they may prove useful in future studies, we collect here explicit expressions for the elastic scattering amplitudes for spin-3/2 particles with potential \eqref{quartics}. In the transversity basis (suppressing an overall factor of $m_{3/2} / \mathcal{M}_{3/2}^3$ and setting $m_{3/2}=m$), 
\begin{widetext}
\begin{align}
& \tilde{\mathcal{A}}^{XX \to XX}_{+3/2,+3/2,+3/2,+3/2} = - \frac{g_7}{16} \frac{ (s-4m^2)^3}{m^4}      \nonumber \\ 
 &- \frac{g_8}{32 m^4 (s-4m^2)} \Bigg[ s^4 + s^3 \left(2 t-8 m^2\right)+2 s^2 \left(16 m^4+20 m^2 t+t^2\right)+32 m^4 \left(8 m^4-4 m^2
   t+t^2\right)   \nonumber \\
  &\qquad\qquad\qquad\qquad+ i \sqrt{stu} \left(64 m^3 \left(t-2 m^2\right)+8 m s^2+16 m s t\right)-16 s \left(8 m^6+10 m^4
   t-3 m^2 t^2\right) \Bigg] ,
  \end{align}
  
  \begin{align}
&\tilde{\mathcal{A}}^{XX \to XX}_{+3/2,+1/2,+3/2,+1/2} = - \frac{g_8}{96 m^4 (s-4m^2)^2} \Bigg[ 
s^5 + s^4 \left(2 t-8 m^2\right)+2 s^3 \left(32 m^4+24 m^2 t+t^2\right) \nonumber \\  &-16 m^2
   s^2 \left(32 m^4+12 m^2 t-3 t^2\right) 
 + i \sqrt{stu} \left(512 m^7+16
   m s^2 \left(2 m^2+t\right)+64 m^3 s \left(t-6 m^2\right)+8 m s^3\right) \nonumber \\
&+32 m^4 s \left(56 m^4-8 m^2 t+t^2\right) +512 m^8 \left(t-4 m^2\right) 
 \Bigg]  - \frac{g_7}{16} \frac{s (s-4m^2)^2 }{4 m^4},
  \end{align}
  
 \begin{align}
  &\tilde{\mathcal{A}}^{XX \to XX}_{+3/2,-1/2,+3/2,-1/2} = - \frac{g_8}{96 m^4 (s-4m^2) } \Bigg[
   s^4 +2 s^3 \left(t-8 m^2\right)+2 s^2 \left(72 m^4-12 m^2 t+t^2\right) \nonumber \\ &-16 m^2 s \left(32 m^4-12
   m^2 t+t^2\right)  
+32 m^4 \left(t-4 m^2\right)^2 + 32 i m^3 s \sqrt{stu} 
   \Bigg]  - \frac{g_7}{16} \frac{ s^2 (s-4m^2)  }{4 m^4},
   \end{align}
    
      \begin{align}
  \tilde{\mathcal{A}}^{XX \to XX}_{+3/2,-3/2,+3/2,-3/2} =& - \frac{g_7}{16} \frac{s^3}{m^4} - \frac{g_8}{32 m^4} \left[  
s^3 + s^2 \left(2 t-8 m^2\right)+2 s \left(t-4 m^2\right)^2 
 \right]  ,
 \end{align}

\begin{align}
& \tilde{\mathcal{A}}^{XX \to XX}_{+1/2,+1/2,+1/2,+1/2} = - \frac{g_7}{16} \frac{ (s-4m^2) (s-12 m^2 )^2 }{9 m^4} 
 - \frac{g_8}{288 (s-4m^2)} \Bigg[   
 s^4 + s^3 \left(2 t-24 m^2\right) \nonumber \\ &-16 m^2 s \left(22 m^2-3 t\right) \left(4 m^2+t\right) +2 s^2 \left(144
   m^4+20 m^2 t+t^2\right) + 32 m^4 \left(72 m^4-4 m^2 t+t^2\right) \nonumber \\ &+ i \sqrt{stu} \left(64 m^3 \left(t-2
   m^2\right)+8 m s^2+16 m s t\right)
 \Bigg]  ,
\end{align} 
 \begin{align}
 \tilde{\mathcal{A}}^{XX \to XX}_{+1/2,-1/2,+1/2,-1/2} =& - \frac{g_7}{16} \frac{ s (s+8m^2)^2 }{9 m^4} - \frac{g_8}{288 m^4} \left[ s^3 + 2 s^2 \left(4 m^2+t\right)+2 s \left(48 m^4+8 m^2 t+t^2\right)  \right]   ,\label{Texplicit} 
\end{align}
\end{widetext}
together with their CPT conjugates, $\mathcal{A}_{\tau_1 \tau_2 \tau_3 \tau_4} = \mathcal{A}_{-\tau_1 -\tau_2 -\tau_3 -\tau_4}^* $ and crossing images, $\mathcal{A}_{\tau_1 \tau_2 \tau_1 \tau_2} = \mathcal{A}_{\tau_2 \tau_1 \tau_2 \tau_1}$
The amplitudes for the other channel, $XY \to XY$, were also computed explicitly, and we confirmed that they indeed satisfy \eqref{Tcrossing} with our phase convention for \eqref{Xpols} and \eqref{Ypols}. 


\end{document}